\documentclass[twocolumn]{aastex63}
\usepackage{CJK}
\usepackage{enumitem}
\usepackage{verbatim}
\usepackage{amsmath}
\usepackage[verbose]{placeins}
\newcommand{\XMM}{ XMM-{\em Newton}}

\shorttitle{XMM-RM catalog}
\shortauthors{Liu et al.}

\begin{document}
\begin{CJK*}{UTF8}{gkai}

  \title{The Sloan Digital Sky Survey Reverberation Mapping Project:
    the XMM-Newton X-ray source catalog and multi-band counterparts}

\author{Teng Liu (刘腾)}
\email{lewtonstein@gmail.com}
\affiliation{Max-Planck-Institut f\"ur extraterrestrische Physik, Giessenbachstra{\ss}e 1, D-85748 Garching bei M\"unchen, Germany}
\author{Andrea Merloni}
\affiliation{Max-Planck-Institut f\"ur extraterrestrische Physik, Giessenbachstra{\ss}e 1, D-85748 Garching bei M\"unchen, Germany}
\author{Torben Simm}
\affiliation{Max-Planck-Institut f\"ur extraterrestrische Physik, Giessenbachstra{\ss}e 1, D-85748 Garching bei M\"unchen, Germany}
\author{Paul J. Green}
\affiliation{Harvard-Smithsonian Center for Astrophysics, MS-4 60 Garden St., Cambridge, MA 02138}
\author{William N. Brandt}
\affiliation{Department of Astronomy \& Astrophysics, 525 Davey Lab, The Pennsylvania State University, University Park, PA 16802, USA}
\affiliation{Institute for Gravitation and the Cosmos, The Pennsylvania State University, University Park, PA 16802, USA}
\affiliation{Department of Physics, 104 Davey Lab, The Pennsylvania State University, University Park, PA 16802, USA}
\author{Donald P. Schneider}
\affiliation{Department of Astronomy \& Astrophysics, 525 Davey Lab, The Pennsylvania State University, University Park, PA 16802, USA}
\affiliation{Institute for Gravitation and the Cosmos, The Pennsylvania State University, University Park, PA 16802, USA}
\author{Tom Dwelly}
\affiliation{Max-Planck-Institut f\"ur extraterrestrische Physik, Giessenbachstra{\ss}e 1, D-85748 Garching bei M\"unchen, Germany}
\author{Mara Salvato}
\affiliation{Max-Planck-Institut f\"ur extraterrestrische Physik, Giessenbachstra{\ss}e 1, D-85748 Garching bei M\"unchen, Germany}
\author{Johannes Buchner}
\affiliation{Max-Planck-Institut f\"ur extraterrestrische Physik, Giessenbachstra{\ss}e 1, D-85748 Garching bei M\"unchen, Germany}
\author{Yue Shen}
\affiliation{Department of Astronomy, University of Illinois at Urbana-Champaign, Urbana, IL 61801, USA}
\affiliation{National Center for Supercomputing Applications, University of Illinois at Urbana-Champaign, Urbana, IL 61801, USA}
\author{Kirpal Nandra}
\affiliation{Max-Planck-Institut f\"ur extraterrestrische Physik, Giessenbachstra{\ss}e 1, D-85748 Garching bei M\"unchen, Germany}
\author{Antonis Georgakakis}
\affiliation{Institute for Astronomy and Astrophysics, National Observatory of Athens, V. Paulou \& I. Metaxa, 11532, Greece}
\author{Luis C. Ho}
\affiliation{The Kavli Institute for Astronomy and Astrophysics, Peking University, 5 Yiheyuan Road, Haidian District, Beijing, 100871, China}
\affiliation{Department of Astronomy, Peking University, 5 Yiheyuan Road, Haidian District, Beijing, 100871, China}

\begin{abstract}
  The XMM-RM project was designed to provide X-ray coverage of the Sloan Digital Sky Survey Reverberation Mapping (SDSS-RM) field.
  $41$ \XMM{} exposures, placed surrounding the Chandra AEGIS field, were taken, covering an area of 6.13 deg$^2$ and reaching a nominal exposure depth of $\sim 15$ ks.
  We present an X-ray catalog of $3553$ sources detected in these data, using a PSF-fitting algorithm and a sample selection threshold that produces a $\sim 5\%$ fraction of spurious sources.
  In addition to the PSF-fitting likelihood, we calculate a second source reliability measure based on Poisson theory using source and background counts within an aperture. Using the Poissonian likelihood, we select a sub-sample with a high purity and find that it has similar number count profiles to previous X-ray surveys.
  The Bayesian method ``NWAY'' was employed to identify counterparts of the X-ray sources from the optical Legacy and the IR unWISE catalogs, using a 2-dimensional unWISE magnitude-color prior created from optical/IR counterparts of Chandra X-ray sources.
  A significant number of the optical/IR counterparts correspond to sources with low detection likelihoods, proving the value of retaining the low-likelihood detections in the catalog.
  $932$ of the XMM-RM sources are covered by SDSS spectroscopic observations. 89\% of them are classified as AGN, and 71\% of these AGN are in the SDSS-RM quasar catalog.
    Among the SDSS-RM quasars, $80\%$ are detectable at the depth of the XMM observations.
  
\end{abstract}

\keywords{
Active galactic nuclei --
Astronomy data analysis --
Quasars --
Supermassive black holes --
X-ray active galactic nuclei --
X-ray astronomy --
X-ray point sources}

\section{Introduction} \label{sec:intro}

The Sloan Digital Sky Survey Reverberation Mapping (SDSS-RM) project \citep{Shen2015,Shen2019} is the
first multi-object RM program with the aim of measuring the black hole masses of a
large representative quasar sample at cosmological distances. This project focuses on a single 7~$\deg^2$
field centered around 213.7$\deg$, 53.1$\deg$ (J2000), in which 849 broad-line quasars with $i_{\mathrm{psf}}\left(\mathrm{SDSS}\right)<21.7$ mag and $0.1<z<4.5$ were selected and continuously monitored with the BOSS spectrographs \citep{Smee2013}, first within the SDSS-III \citep{Eisenstein2011,Dawson2013} and then within the SDSS-IV \citep{Blaton2017} programs, forming a unique sample of quasars with unprecedented multi-band, multi-epoch imaging and spectroscopy.
The field is fully covered by the Panoramic Survey Telescope and Rapid Response System (Pan-STARRS) Medium Deep Field (MDF) survey \citep{Kaiser2010,Tonry2012}, the WISE survey \citep{Wright2010,Lang2016}, and the Faint Images of the Radio Sky at Twenty cm (FIRST) radio survey \citep{White1997}.
It is also covered partially by the Galaxy Evolution Explorer (GALEX) near-ultraviolet (NUV) survey \citep{Gezari2013}, United Kingdom Infrared Telescope (UKIRT) near-infrared imaging, and the All-wavelength Extended Groth strip International Survey (AEGIS) \citep{Davis2007}.
Since the X-ray band is of particular importance to the study of AGN, as it provides important information about the AGN environment near the black hole, we initiated the XMM-RM project to survey this field in the X-ray band with \XMM{}.
This paper presents the X-ray catalog and optical/IR counterparts of the X-ray sources in this field.
The WMAP cosmology with $\rm \Omega_{m}$= 0.272, $\rm \Omega_{\Lambda}$ = 0.728 and $H_{0}$ = 70.4 km $\rm s^{-1}$ $\rm Mpc^{-1}$ \citep{Komatsu2011} is adopted.

\section{Observations and data reduction}
  
\subsection{The data}
\label{sec:xmmobs}

\begin{figure}[htbp]
\epsscale{1}
\plotone{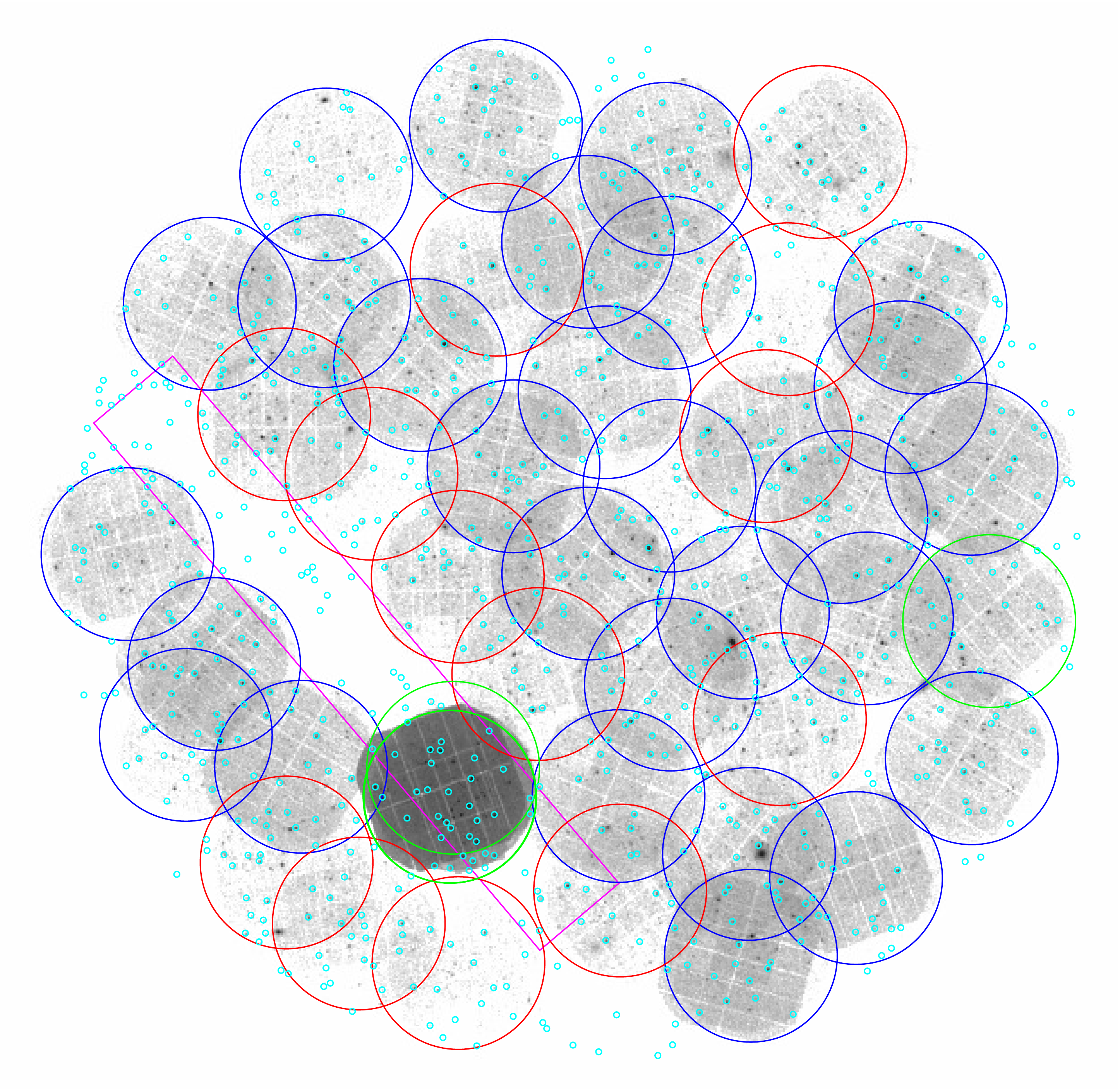}
\caption{
The X-ray (0.5-7.5~keV) mosaic image of all the XMM observations with XMM pointings (30\arcmin~diameter circles) and RM-quasar positions (cyan points).
Red and blue circles indicate the $13+28$ pointings from AO15 and AO16, respectively.
Green circles correspond to existing archival XMM observations.
The magenta box is the Chandra AEGIS/EGS field.
}
\label{fig:observations}
\end{figure}

The \XMM{} observations cover a total sky area of $\sim 6.13$\,$\mathrm{deg^{2}}$ within the 7\,$\mathrm{deg^{2}}$ RM field. The main
data set is formed by 41 XMM observations with individual exposure times of
$\sim$15~ks, with 13 pointings obtained in AO15 (PI P. Green) and 28 in AO16 (PI A. Merloni).
All observations were taken in full frame mode of the EPIC-PN/MOS cameras.
Most observations were performed with the thin filter; only four observations
required the medium filter due to nearby bright field stars.
The footprint of the XMM-RM survey is shown in Fig. \ref{fig:observations}.
The pointings have a regular 22\arcmin\, spacing and overlap sufficiently to
provide a uniform coverage of sensitivity.

In addition to these XMM-RM dedicated observations, we also reanalyze seven archival XMM
observations in the RM field (see Fig.~\ref{fig:observations}), including five 
Groth-Westphal Strip observations (PI R. Griffiths) and two object-targeted
observations (0503960101, PI M. Agueros; 0723860101, PI D. Lin).
Two observations (0127920401, 0127920901) are excluded because of high background flaring during the exposures.
In total, 90\% (756/849) of the SDSS-RM quasars are covered by the XMM observations.
Table~\ref{tab:xmmobs} lists the details of the XMM observations presented in this work.

\begin{table*}
\caption{XMM-RM observations}
\centering
\begin{tabular}{ccccllll}
\hline\hline
ObsID & RA & DEC & DATE & PN MODE & PN $T_{\mathrm{EXP}}$ & MOS MODE & MOS $T_{\mathrm{EXP}}$ \\
& [deg] & [deg] & & & [ks (\%)] & & [ks (\%)] \\
\hline
0127921001 & 214.3060 & 52.3726 & 2000-07-21 &	EFF-Thin & 45(87) & FF-Thin & 53.1(95)\\
0127921201 & 214.3045 & 52.3725 & 2000-07-23 &	EFF-Thin & 14.4(98) & FF-Thin & 18.4(99)\\
0127921101 & 214.3044 & 52.3725 & 2000-07-23 &	EFF-Thin & 3.6(97) & FF-Thin & 7.5(100)\\
0503960101 & 211.7222 & 52.8577 & 2007-06-21 &	FF-Thin & 12.7(53) & FF-Thin & 18(68)\\
0723860101 & 214.2908 & 52.4564 & 2014-01-05 &	 &  & FF-Thin & 20.3(71)\\
0765070301 & 212.4808 & 54.2289 & 2015-12-24 &	FF-Thin & 13.2(83) & FF-Thin & 15.8(90)\\
0762500301 & 213.5022 & 52.1000 & 2015-12-26 &	FF-Thin & 10.7(72) & FF-Thin & 13.7(83)\\
0762500201 & 214.6906 & 53.3064 & 2015-12-26 &	FF-Thin & 2.8(18) & FF-Thin & 6.4(37)\\
0762500901 & 214.2633 & 51.8914 & 2016-01-13 &	FF-Thin & 3.4(21) & FF-Thin & 8.4(48)\\
0762500501 & 213.8848 & 52.7277 & 2016-01-15 &	FF-Thin & 4.3(27) & FF-Thin & 13.3(76)\\
0765080101 & 214.0848 & 53.8985 & 2016-01-21 &	FF-Thin & 5(25) & FF-Thin & 11.7(52)\\
0765080801 & 215.0757 & 52.1784 & 2016-01-23 &	FF-Medium & 5.2(31) & FF-Medium & 8.4(45)\\
0765090801 & 212.6571 & 53.7758 & 2016-01-23 &	FF-Thin & 4.5(30) & FF-Thin & 6.7(40)\\
0765081001 & 212.7749 & 53.4096 & 2016-01-27 &	FF-Thin & 11.1(79) & FF-Medium & 15.4(99)\\
0765080601 & 212.7343 & 52.5897 & 2016-01-27 &	FF-Thin & 10.4(95) & FF-Thin & 12.6(100)\\
0765080901 & 214.7323 & 52.0041 & 2016-01-27 &	FF-Thin & 3.2(29) & FF-Medium & 12.6(100)\\
0762500401 & 214.2715 & 53.0097 & 2016-01-31 &	FF-Thin & 10.6(66) & FF-Thin & 15.5(88)\\
0762500101 & 215.1179 & 53.4753 & 2016-02-02 &	FF-Thin & 12.1(95) & FF-Thin & 14.3(99)\\
0804270101 & 214.0023 & 53.3293 & 2017-05-09 &	FF-Thin & 17.4(92) & FF-Thin & 20.6(100)\\
0804270201 & 214.0879 & 54.3156 & 2017-05-10 &	FF-Thin & 13.7(98) & FF-Thin & 15.6(100)\\
0804270601 & 212.8912 & 51.9066 & 2017-05-11 &	FF-Thin & 9.1(48) & FF-Thin & 20.6(100)\\
0804270401 & 212.0070 & 53.7704 & 2017-05-19 &	FF-Thin & 15.5(82) & FF-Thin & 17(83)\\
0804270501 & 213.5061 & 52.3733 & 2017-05-20 &	FF-Thin & 11(79) & FF-Thin & 15.3(98)\\
0804270301 & 213.2358 & 53.8581 & 2017-05-24 &	FF-Thin & 12.7(98) & FF-Thin & 13.2(100)\\
0804270701 & 215.4901 & 53.7919 & 2017-05-24 &	FF-Thin & 13.7(99) & FF-Thin & 15.5(100)\\
0804270901 & 215.0135 & 52.4560 & 2017-05-29 &	FF-Thin & 18.1(95) & FF-Thin & 20.6(100)\\
0804270801 & 211.8265 & 52.4625 & 2017-05-30 &	FF-Thin & 14.8(99) & FF-Thin & 16.6(100)\\
0804271301 & 213.6429 & 53.0179 & 2017-06-08 &	FF-Thin & 18.5(97) & FF-Thin & 20.6(100)\\
0804271401 & 213.2518 & 52.6945 & 2017-06-09 &	FF-Thin & 13.5(96) & FF-Medium & 15.4(99)\\
0804271501 & 212.8926 & 52.2013 & 2017-06-09 &	FF-Thin & 13.5(96) & FF-Thin & 15.1(97)\\
0804271001 & 215.4357 & 52.7492 & 2017-06-12 &	FF-Thin & 17.9(94) & FF-Thin & 20.6(100)\\
0804271201 & 214.4570 & 53.6221 & 2017-06-13 &	FF-Thin & 12.8(91) & FF-Thin & 15.6(100)\\
0804271101 & 214.9297 & 53.8047 & 2017-06-13 &	FF-Thin & 13.4(96) & FF-Thin & 15.6(100)\\
0804271601 & 212.3896 & 52.1253 & 2017-07-14 &	FF-Thin & 7.5(41) & FF-Thin & 14.8(74)\\
0804271901 & 212.9022 & 52.8998 & 2017-07-15 &	FF-Thin & 13.2(94) & FF-Thin & 15.6(100)\\
0804271701 & 213.5579 & 53.5425 & 2017-07-20 &	FF-Thin & 8.2(49) & FF-Thin & 13.6(72)\\
0804272101 & 213.6338 & 53.9783 & 2017-07-22 &	FF-Thin & 4.4(24) & FF-Thin & 5.7(34)\\
0804272301 & 215.8629 & 53.0618 & 2017-07-23 &	FF-Thin & 11(79) & FF-Thin & 15.5(99)\\
0804272501 & 213.2493 & 54.1879 & 2017-07-23 &	FF-Thin & 13.5(96) & FF-Thin & 15.6(100)\\
0804272801 & 214.9270 & 54.1715 & 2017-07-24 &	FF-Thin & 6.7(38) & FF-Thin & 8(43)\\
0804271801 & 213.2472 & 53.2703 & 2017-11-09 &	FF-Thin & 6(45) & FF-Thin & 6.4(56)\\
0804272201 & 212.3076 & 52.8757 & 2017-11-15 &	FF-Thin & 13.3(95) & FF-Thin & 15.3(98)\\
0804272001 & 215.5636 & 52.5425 & 2017-11-16 &	FF-Thin & 10.4(72) & FF-Thin & 4.1(91)\\
0804272401 & 212.4188 & 53.1708 & 2017-11-27 &	FF-Thin & 12.3(88) & FF-Thin & 14.9(96)\\
0804272601 & 212.1157 & 53.5421 & 2017-11-27 &	FF-Thin & 13.6(97) & FF-Thin & 15.4(99)\\
0804272701 & 211.7837 & 53.2997 & 2017-12-03 &	FF-Thin & 18.7(98) & FF-Thin & 20.7(100)\\
\hline
\end{tabular}
\tablecomments{Column 1: XMM observation ID; columns 2--3: aim point (J2000); column 4: observation date ; columns 5,7: observation mode and filter for PN and MOS (FF for PrimeFullWindow and EFF for PrimeFullWindowExtended) ; columns 6,8: cleaned exposure time (``ONTIME'') after removal of background flares and its ratio to raw exposure time in percentage terms.}
\label{tab:xmmobs}
\end{table*}

The Extended Groth Strip (EGS, $\sim 0.67 \deg^2$), which is located within the SDSS-RM field (see Fig. \ref{fig:observations}), was observed by Chandra. The entire EGS is covered to a depth of 200\,ks, while the central 0.29\,$\mathrm{deg^{2}}$ reaches a nominal depth of 800\,ks as part of the AEGIS-X Deep survey. These observations are described in \citet{Nandra2005,Laird2009,Goulding2012} and \citet{Nandra2015}, with the X-ray source catalogs and the corresponding multi-band counterparts presented.
This paper focuses only on the XMM data.
  
\subsection{Data processing}
\label{sec:srcdet}

The data processing was performed with the \XMM{} Science Analysis Software (SAS)
version 18.0.0 \citep{Gabriel2004}. Aiming to optimize the sensitivity for faint point sources, we developed a dedicated data reduction strategy to
suppress noise.

The first step is the creation of EPIC PN and MOS event files from the Observation Data Files (ODF) executing the \texttt{EPCHAIN} and \texttt{EMCHAIN} SAS tasks. In addition to the standard event file, we also create an
out-of-time (OOT) event file for the PN data.
To avoid spurious detections at the CCD edges, we remove the pixels along the edges of
the PN and MOS CCDs by flagging them as dead pixels.
For the PN, we exclude events with patterns larger than $4$ or with energies in the ranges of
instrument lines \citep{Ranalli2015}, i.e., 1.39-1.55 (Al) and
7.35-7.60, 7.84-8.28, 8.54-9.00 keV (Cu).
We adopt a stricter event flag filter than the commonly-used ``XMMEA\_EP'',
excluding the ``CLOSE\_TO\_CCD\_BORDER'', ``CLOSE\_TO\_CCD\_WINDOW'', ``OUT\_OF\_FOV'', and ``OUT\_OF\_CCD\_WINDOW'' events (flag code 0xefb0006). At energies below 1 keV, we also reject
``ON\_OFFSET\_COLUMN'' events (0xefb000e).
For the MOS, we exclude events with patterns larger than $12$ or with energies in the ranges of
1.39-1.55 (Al), 1.69-1.80 keV(Si). In addition to the commonly-used ``XMMEA\_EM''
filter, we also exclude the ``CLOSE\_TO\_CCD\_BORDER'' and ``CLOSE\_TO\_CCD\_WINDOW''
events (0x766ba006).

We adopt a two-step procedure to reject background flares.
First, strong flares are excluded using the \texttt{espfilt} ``ratio'' method in the 8-12 keV band, allowing a count rate ratio of inside to outside field-of-view (FOV) of 1.5.
Excluding such flares, images and exposure maps are constructed in the 0.5-7.5 keV band, and sources are detected using \texttt{eboxdetect}.
Then, a lightcurve is extracted from the source-free region in the 0.5-7.5 keV band, which is the band used for source detection in this work.
We use \texttt{bkgoptrate} to find the count rate threshold at which
the maximum signal-to-noise (S/N) ratio is achieved after excluding the bins
above the threshold. This threshold defines cleaner good-time-intervals (GTIs).
Finally, we merge all GTIs generated by \texttt{ep/emchain} for each CCD and the GTIs from \texttt{espfilt} and from \texttt{bkgoptrate} using the ``AND'' mode.
The MOS1 and MOS2 lightcurves are merged before running
\texttt{bkgoptrate}, and only one final GTI was constructed for MOS1 and MOS2.
In Table~\ref{tab:xmmobs}, the cleaned exposure times are listed and also expressed as percentages of the raw exposure time.

\subsection{Astrometric correction}
Initially, we run source detection as described in \S~\ref{sec:det} in the full band without any astrometric correction.
For each observation, we compare the detected sources with an optical/IR reference catalog created by \citet{Rosen2016} based on the SDSS (Abazajian et al. 2009), 2MASS (Skrutskie et al. 2006), and USNO-B1.0 (Monet et al. 2003) catalogs, using the task \texttt{catcorr}, which calculates both the shift and rotation corrections and the corresponding errors. The shift is typically $<0.5\arcsec$.
We apply the corrections to the attitude file, and reconstruct the event files using the task \texttt{evproject}.
The analyses hereafter are based on these astrometrically-corrected event files.
After source detection, we convert the errors of the corrections into an additional systematic error for each source (the ``SYSERRCC'' column in the catalog) according to its off-axis angle.
In the final catalog, one source could be detected in multiple observations which contribute different systematic errors. We choose the largest one for simplicity.

\section{Source Detection}
\label{sec:det}
\subsection{Images}
Three energy bands are used for source detection: full (0.5-7.5 keV), soft (0.5-2 keV), and hard (2-7.5 keV).
A pixel scale of 4$\arcsec$ is used.
In order to improve the measurement of the background, which is essential to the
detection of faint sources, we also create Filter-wheel-closed
(FWC) background maps and OOT maps (for PN only, also hereafter whenever OOT is mentioned) for each observation using the
\texttt{eimageget} task.
To increase the S/N of the FWC map while taking account of the
long-term instrument variability, for each observing mode we divide all the FWC
observations obtained from 2001 to 2017 into five epochs, each one having
approximately the same summed exposure time, and merge the events in each epoch.
We find that stacked FWC map is highly flat and does not have a sufficient S/N to reveal any feature at the small, PSF scales.
Since in this work we are particularly interested in faint point
sources rather than extended emission on large scales, we calculate the mean
value of the FWC map at each epoch. These values are used in the subsequent analysis.

To construct the OOT map for PN, the OOT image is smoothed only along the CCD reading columns (DETY direction in detector coordinates), making a stripe-pattern image which assigns the same value to all the pixels of each CCD with the same DETX.

We make the exposure maps using the \texttt{expmap} task with the highest
positional accuracy (0.02$\arcsec$) in order to match the sky coordinates of the
images exactly.

\subsection{Background maps}
We detect sources in each image by running \texttt{eboxdetect} twice with a box size
of 5 pixels and a minimum likelihood of 8: the first iteration in local mode (without background map), and the second using the background map generated by \texttt{esplinemap} on the basis of the catalog detected in the first run.
We exclude the detected sources using \texttt{esplinemap}.
The removal of the source signal does not have to be perfect, because the residual source remnants can be excluded later through sigma clipping.

The source-removed image is adaptively smoothed to achieve a S/N of 10 using \texttt{asmooth}.
The different convolvers used in different parts of the image are saved.
The corresponding exposure map and OOT map are smoothed using the same convolvers, keeping them consistent with each other.

Subtracting the FWC flux and the smoothed OOT map from the smoothed image produces the X-ray background component which, in principle, follows the vignetted exposure map, i.e., the flux map generated by dividing the residual image by the
vignetted exposure map should be flat.
However, the vignetted exposure maps might not be perfectly accurate because: 1) the vignetting maps, which are energy-dependent, are generated at single typical energies of each band; 2) the soft-proton background component, which does not follow the same vignetting map as the X-ray photons, cannot be completely removed.
Deviations will appear as a centrosymmetric pattern.
Therefore, we model the flux map with a centrosymmetric model -- a 1-dimensional quadratic smoothing spline model as a function of off-axis angle
\footnote{We make use of the \texttt{interpolate.LSQUnivariateSpline} function in the
  ``scipy'' package.}, and calculate this spline with the values of all the pixels.
By iterative 3-$\sigma$ clipping, the potential remnants of source signals can be excluded.
Finally, the background map is generated by adding the FWC flux and OOT map back
to the spline-modeled background map.

Our background map is generated on the assumption that the X-ray background has a uniform flux across the FOV. This assumption can be violated if a large extended X-ray source falls inside or nearby the FOV, but there is no such case in this field.

\subsection{Simultaneous PSF-fitting Detection}
It is common practice in XMM surveys to have large overlaps between nearby pointings, and ours is no exception.
In overlapping regions, we have the possibility to increase 
the detection sensitivity. However, the varying PSF across the EPIC FOV makes it a bad idea to stack the images, since the detailed PSF shapes, which differ in different observations, are lost during the stacking.
In this work, we detect sources through simultaneous PSF fitting using the task \texttt{emldetect} on the images of
each camera in each observation without stacking. In this approach, not only the
data, but also the PSF shapes (``ellbeta'' model) at each position in each observation are fully exploited.

Due to limitations in the capabilities of \texttt{emldetect}, some special treatment of the images is needed. To avoid a large number of 
images input to \texttt{emldetect}, we use a 4x4 grid to divide the entire field into smaller cells, each of which
is covered by a few tens of images at most. For each cell, we create an image for
each camera in each observation using the same frame of the cell, adding a
36$\arcsec$ padding region around the edges. The exposure maps, background maps, and detection masks are
reprojected to the same frame using the \texttt{CIAO} task \texttt{reproject\_image}.
We run \texttt{emldetect} on these images, producing a catalog for each cell.
When we merge the catalogs of all the cells, the padding (overlapping) regions are checked in order to avoid missing any source which is detected as being just outside the cell in all cases and to avoid any duplicate source which is detected by more than one cell as being inside the cell.

The PSF fitting is done within a radius of 5 pixels (20$\arcsec$). Multi-source fitting is
applied for sources located within 10 pixels of each other.
The detection mask is chosen as the region where the full-band exposure time is
at least 10\% of the maximum in the FOV.

The PSF fitting provides a detection likelihood $L$, which is equivalent to the probability $p$ for a random Poissonian fluctuation to have caused the observed source counts in terms of $L=-\ln(p)$.
Note that for small numbers of source photon counts,  this relation is only a rough estimation \citep{Cash1979}.
We selected a relatively low threshold of $L>3$, which corresponds to a nominal spurious fraction of 5\%.

\subsection{Independent detections in three bands}
Considering the variety of spectral shapes of X-ray sources, simply using the information in the full band does not always produce the best sensitivity. For very soft or very hard sources, using
the data in only the soft or hard band could result in a higher S/N ratio.
To account for this we ran source detections independently in the full, soft, and hard bands and
then merged the catalogs.

We match the sources detected in different bands as follows. First, the output sources of the PSF fitting in different bands triggered by the same input source are considered as one. Second, we allow a minimum source separation of 9$\arcsec$, which is approximately the half-energy PSF radius of XMM EPIC at 1.5 keV. Taking the positional error into account, when the separation of two sources is less than $9\arcsec+\sqrt{\sigma_{Source 1}^2 + \sigma_{Source 2}^2}$, they are interpreted as a single object.
This threshold will not likely cause any mis-matches, considering that the minimum separation between full-band detected sources is $\sim 12\arcsec$.

\subsection{Three-pass detections}
Before running the PSF fitting software, we require an initial set of positions of candidate sources. For each cell, we stack
all the images, exposure maps, and background maps, and feed them to the
\texttt{CIAO} task \texttt{wavdetect}, using wavelet scales of 2,3,4,6,8,16
pixels and a threshold of $10^{-4}$. This procedure results in a large seed catalog with
many spurious sources. We merge the catalogs detected in the three bands and input
the merged catalog to the next pass, \texttt{emldetect}.

Running PSF fitting on the raw catalog from \texttt{wavdetect}, we compile a catalog and select sources with a detection likelihood $>3$.
This new list of sources is fed back to \texttt{emldetect} to repeat the PSF fitting. 
Since a large number of unrealistic, faint sources are removed from the input list, un-necessary multi-PSF fittings are prevented.
Based on the second iteration of PSF fitting, we again select the sources with a detection likelihood $>3$, then merge the catalogs detected in each cell, and in a last step merge the catalogs in each energy band.
The final catalog contains $3553$ sources. It is publicly available along with this paper and described in Appendix~\ref{app:Xraycolumns}.

\subsection{Extended sources}
\label{sec:clusters}
A source extent likelihood is calculated by \texttt{emldetect} as the likelihood difference between a fit of the source surface brightness with a beta model and that with the PSF model.
A minimum extent likelihood of $5$ is adopted.
In the final catalog, there are only $19$ sources with an extent likelihood $>10$. They are considered as extended sources.
We remark that this small sample of extended sources is neither complete nor clean, because the detection and selection in this work are optimized for unresolved sources but not for extended sources. For example, the PSF fitting within a small region is inefficient at identifying diffuse emission; fitting an extended source with the PSF results in residual diffuse emission which may easily be misidentified as a group of faint sources; and bad columns of the CCD could affect the fitting and cause false extent classifications.
A detailed study of galaxy clusters is out of the scope of this paper. We focus on unresolved sources, especially AGN, which consititue a large majority of the X-ray sources in such an extragalacitc field.
We cross-correlate the sources that have an extent $>0$ or have an X-ray neighbor within 1$\arcmin$, with the \citet{Wen2012} SDSS cluster catalog within a distance of 1.5$\arcmin$, and visually inspect the matches. Half ($13$) of the extent $>0$ sources can be associated with SDSS clusters.
We also identify another $24$ unresolved sources which are likely SDSS clusters mis-classified as unresolved or likely substructures of the SDSS clusters.
In total, these sources are attributed to $15$ SDSS clusters. The corresponding cluster ID in the \citet{Wen2012} catalog is added in our catalog (the ``SDSS\_Cluster'' column).

\section{The XMM-RM source catalog}
\label{sec:cat}
\subsection{Average flux}
For each camera, we calculate the Energy Conversion Factors (ECF) from our full, soft, and hard band count rates to 0.5-10, 0.5-2, and 2-10 keV fluxes respectively, using the response files generated at the aimpoint of each camera, assuming an absorbed power-law model with a slope of 1.7 and Galactic absorption \citep{HI4PI}. The narrow instrument-line bands, which are excluded when making the images, are also excluded here.
With these ECFs, we convert the count rate of each source in each camera and each observation to flux.

\begin{figure}[htbp]
\epsscale{1}
\plotone{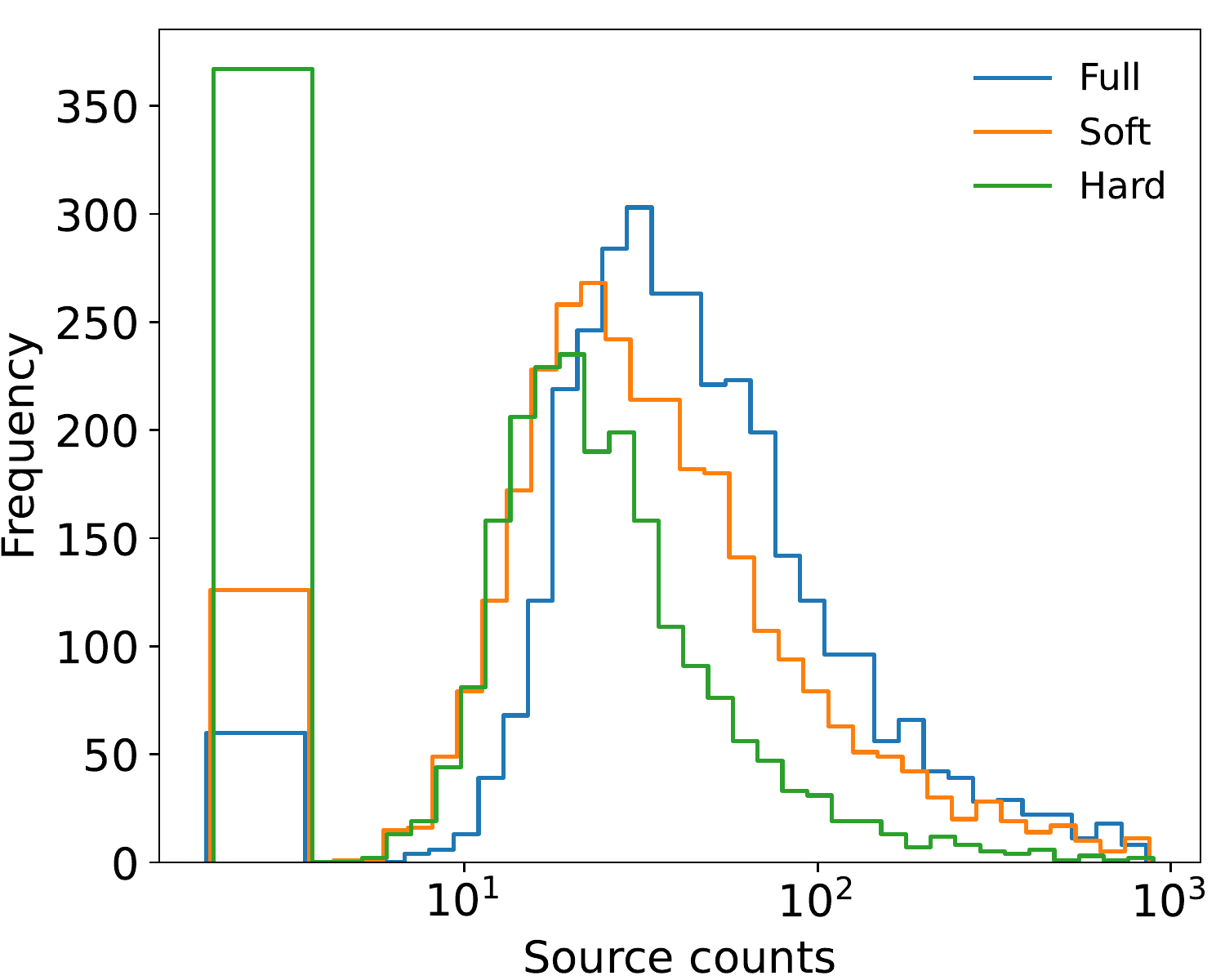}
\plotone{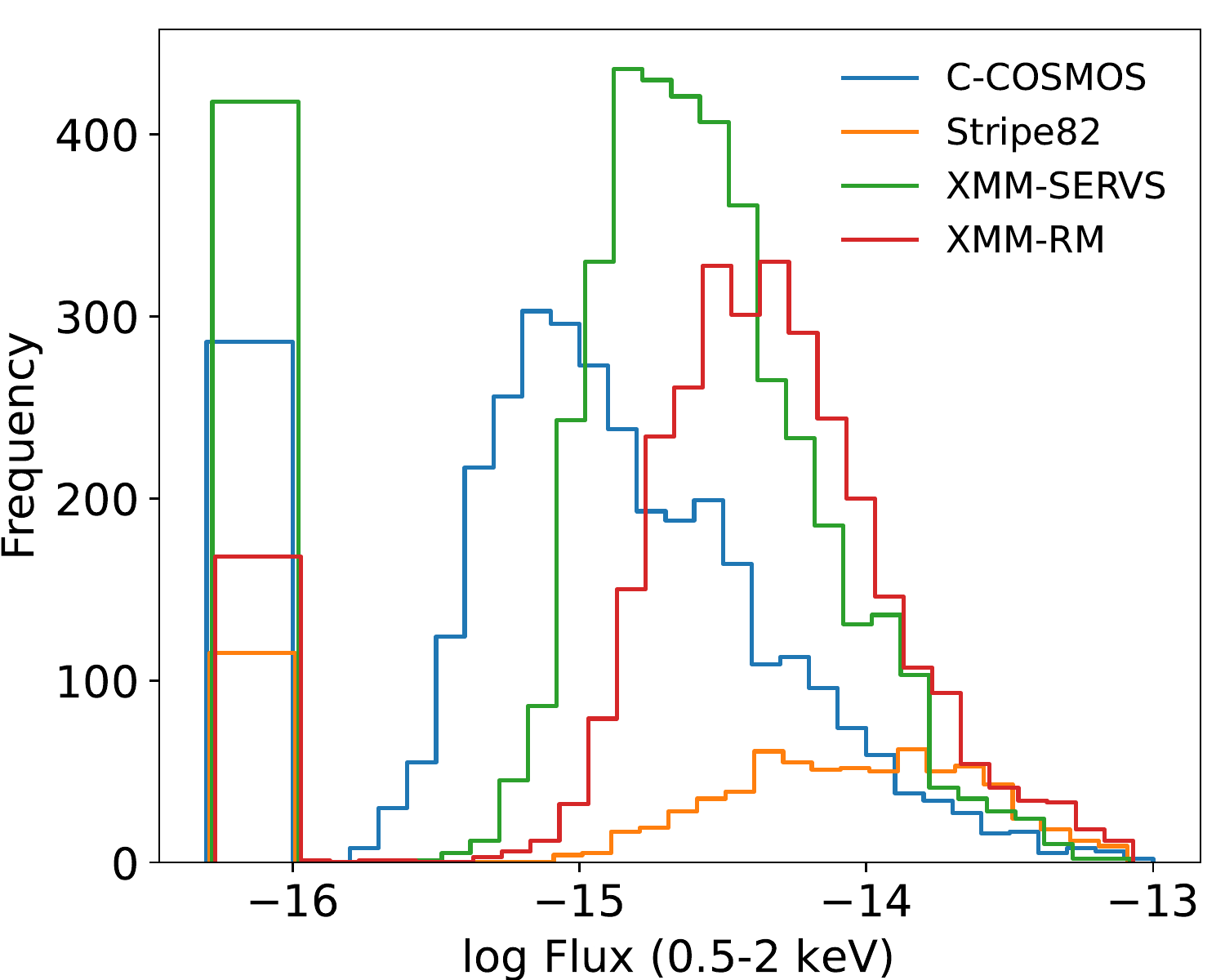}
\caption{
The distributions of source counts in the three bands (upper panel) and the 0.5-2 keV fluxes (lower panel) of the XMM-RM sources in comparison with the C-COSMOS and Stripe 82 Chandra surveys and the XMM-SERVS survey.
Sources that are undetected and thus have 0 counts and flux in a specific band are plotted in special bins at the lower end of the distribution, showing their number in terms of area in the histogram bins.
  }
\label{fig:softflux}
\end{figure}

In each observation, the single-epoch flux of each source is
calculated by averaging among the cameras.
When the PSF weighted on-chip fraction of one source is $<0.8$ in one camera, this detection is excluded from the flux calculation, because such detections often lie at the border of the FOV or in CCD gaps, where the calibration is inaccurate and the flux correction is less reliable.
The average flux of each source is calculated as the exposure-time weighted mean flux among the multiple
observations. 
When the three cameras have different exposure time in one observation, the longest one is used.
A source might have a flux of 0 in one camera during one observation; such cases are not excluded in the flux averaging.
The distribution of soft fluxes is shown in Fig.~\ref{fig:softflux}, in comparison with the flux distributions of the C-COSMOS \citep{Civano2016}, Stripe 82 Chandra \citep{LaMassa2016}, and XMM-SERVS \citep{Chen2018} catalogs.

\subsection{EEF map and ECF map}
\label{sec:EEF_ECF_map}
\begin{figure}[htbp]
\epsscale{1}
\plotone{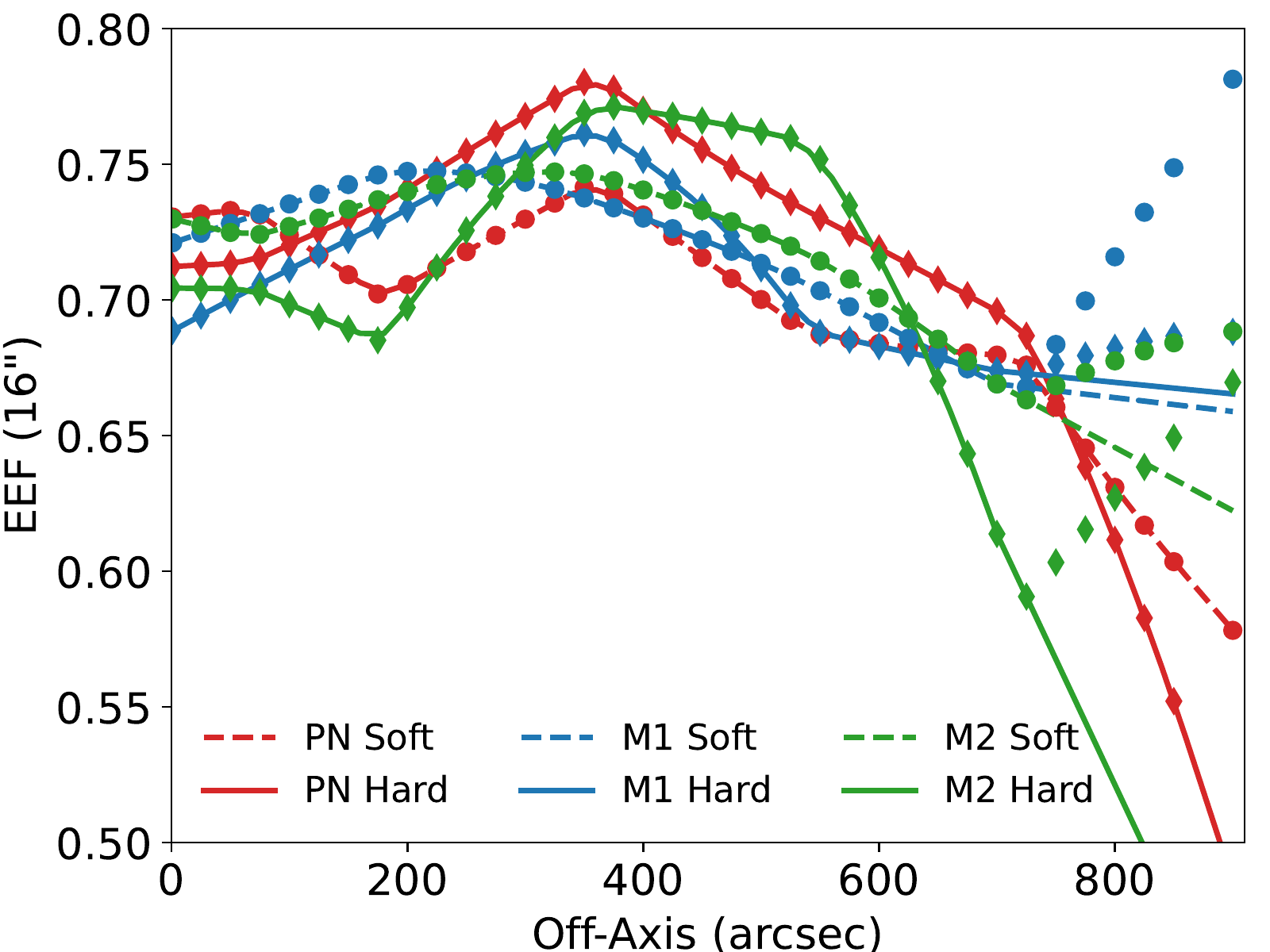}
\caption{16$\arcsec$ Enclosed Energy Fraction (EEF) as a function of off-axis angle. The lines are produced by the spline fitting. The increasing parts at large off-axis angles are excluded from the fitting.
}
\label{fig:EEF_OA}
\end{figure}

For each camera, we generate ``ellbeta'' PSF images using the task
\texttt{psfgen} at a series of off-axis angles and calculate the 16$\arcsec$ Enclosed Energy Fraction (EEF)
for them.
As shown in Fig.~\ref{fig:EEF_OA}, at high off-axis angles, the EEF curve turns over and increases with
off-axis, indicating the PSF model is suspect in such cases.
Thus we exclude this increasing part.
By interpolating and extrapolating the EEF as a function of off-axis angle, we build
an EEF map for each camera in each observation.
These EEF maps are stacked to create an EEF map for the
whole field. The value at each position is calculated as the weighted mean of all the EEF maps covering this position.
For the MOS cameras, we use the exposure maps as the weights; while for PN, we use the exposure
map divided by 0.4, since the effective area of each MOS camera is about $0.4$ of that of PN.

\begin{figure*}[htb]
\epsscale{0.8}
\plotone{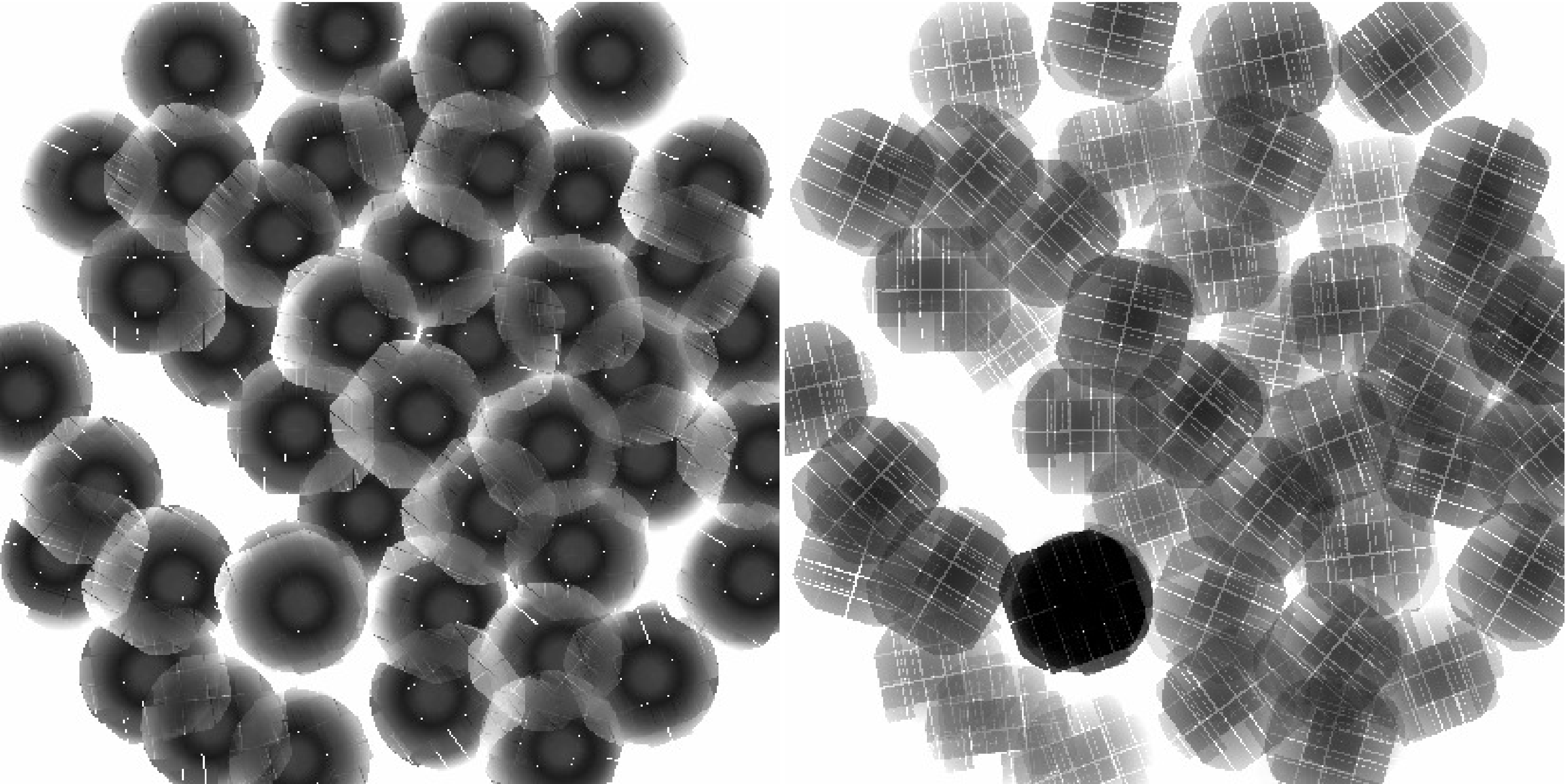}
\caption{Full band stacked Enclosed-Energy-Fraction (EEF) map (left) and Exposure~$\times$~Energy-Conversion-Factor (Exp-ECF) map (right).
  Darker color indicates a higher value.
}
\label{fig:EEF_ECF}
\end{figure*}

For each camera in each observation, we construct an ECF map by filling the FOV
with the ECF of the camera. Each ECF map is multiplied by the corresponding vignetted exposure map, creating an Exp-ECF map. Stacking all of them produces an Exp-ECF map for the entire field, which stores the conversion factor from source net counts to flux at each position.
The stacked full-band EEF map and Exp-ECF map are presented in Fig.~\ref{fig:EEF_ECF}.

\subsection{Poissonian likelihood and sensitivity map}
\label{sec:poissonianlikelihood}
\begin{figure}[htb]
\epsscale{0.9}
\plotone{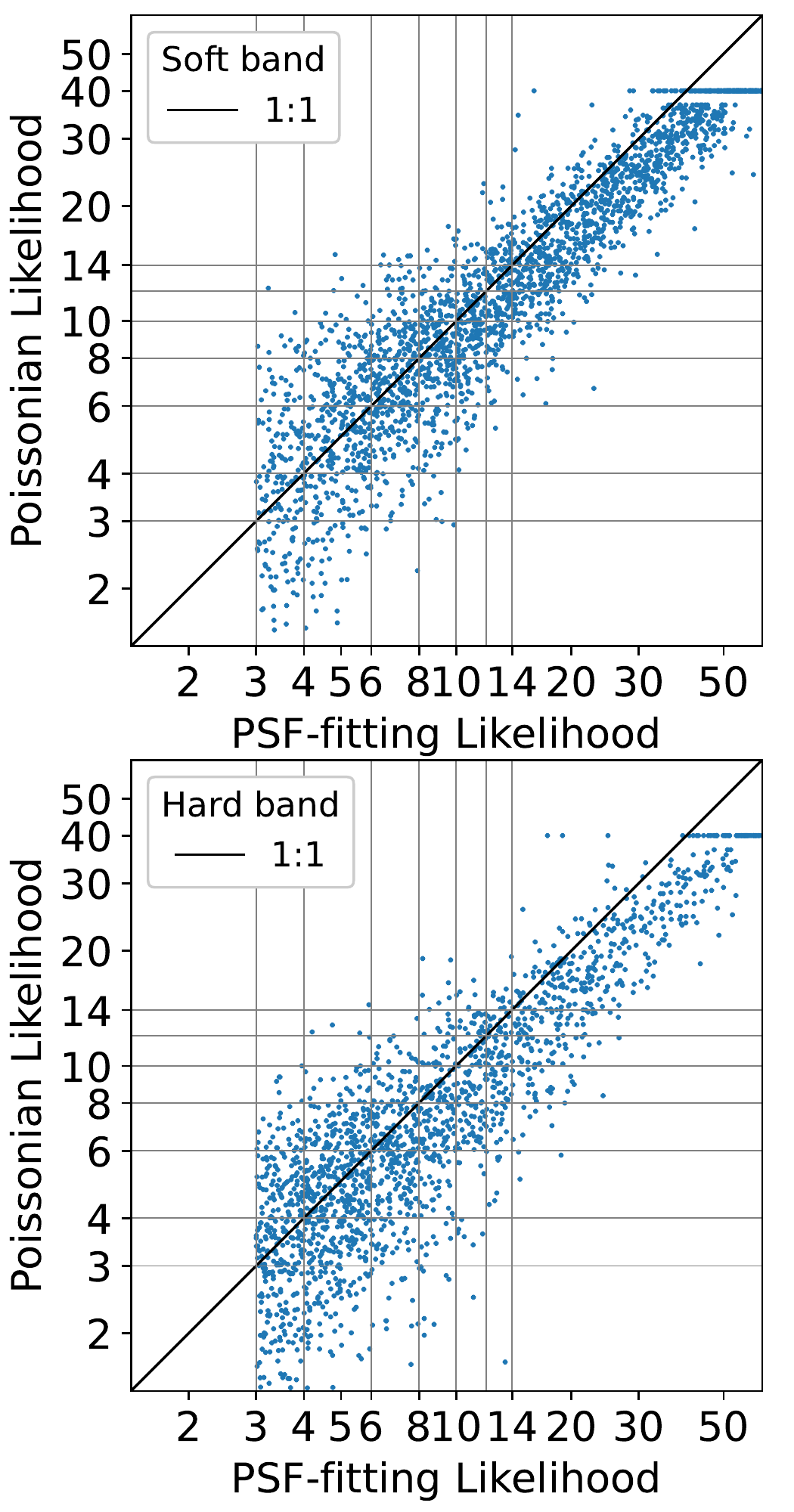}
\caption{
    Comparison between the Poissonian likelihood and the PSF-fitting likelihood in the soft (upper) and hard (lower) band.
  }
\label{fig:DET_ML_AML}
\end{figure}

  In additional to the source detection likelihood measured with PSF-fitting, we calculate the Poissonian likelihood based on aperture source and background counts \citep{Georgakakis2008}.
We apply the task \texttt{eregionanalyses} on the stacked images, background maps, and exposure maps to measure the source and background counts and count rates in a 16$\arcsec$-radius (4 pixels) circular aperture.
For an aperture source counts $S$ and background counts $B$ (note $S+B$ is always an integer), the Poisson probability of the source being spurious is $Poisson_{B}(\geqslant S+B) = \gamma(S+B,B)$, where $\gamma$ is the regularized lower incomplete gamma function, and this probability is converted to a likelihood as $L=-\ln(p)$.
In the case of blending with nearby sources, the aperture source counts are overestimated.
To reject such contamination, when the distance of a source to its nearest neighbor is $<32\arcsec$, we replace the aperture
source counts with that measured from PSF fitting, i.e., the \texttt{emldetect} source counts multiplied by the EEF at the source position.

  The two likelihoods can be significantly different in individual cases, as compared in Fig.~\ref{fig:DET_ML_AML}.
At high likelihoods, the Poissonian likelihood is smaller than the PSF-fitting one.
At low likelihoods, the Poissonian likelihood becomes relatively higher and comparable with the PSF-fitting one, but with a large scatter.
In consideration of the large scatter, we remark that, a PSF-fitting likelihood selected sample can be incomplete with respect to the Poissonian likelihood, as some sources that could survive the Poissonian likelihood threshold might be already removed during the PSF-fitting pre-selection. This is one of the reasons that we choose a PSF-fitting threshold as low as 3 when creating the XMM-RM catalog.

  The Poissonian likelihood is not as good as the PSF-fitting likelihood in distinguishing source signal and fluctuations; however, a Poissonian likelihood threshold can be straightforwardly converted to a map of flux limit in the field, i.e., a sensitivity map.
  We construct an aperture-background-counts map by convolving the background map with a 16$\arcsec$ circular kernel filled with unity value.
  The value of background counts in this map is converted to a minimum source counts in the aperture required to achieve a given likelihood \footnote{We make use of the \texttt{special.gammainccinv} function of the ``scipy'' package.}, creating a map of aperture-source-counts limit ($M$).
  Dividing this map by the EEF map and the Exp-ECF map (Fig.~\ref{fig:EEF_ECF}), it can be converted to flux in units of erg cm$^{-2}$ s$^{-1}$.

\subsection{Sky coverage and number counts}
\label{sec:logNlogS}
\begin{figure}[htb]
\epsscale{1}
\plotone{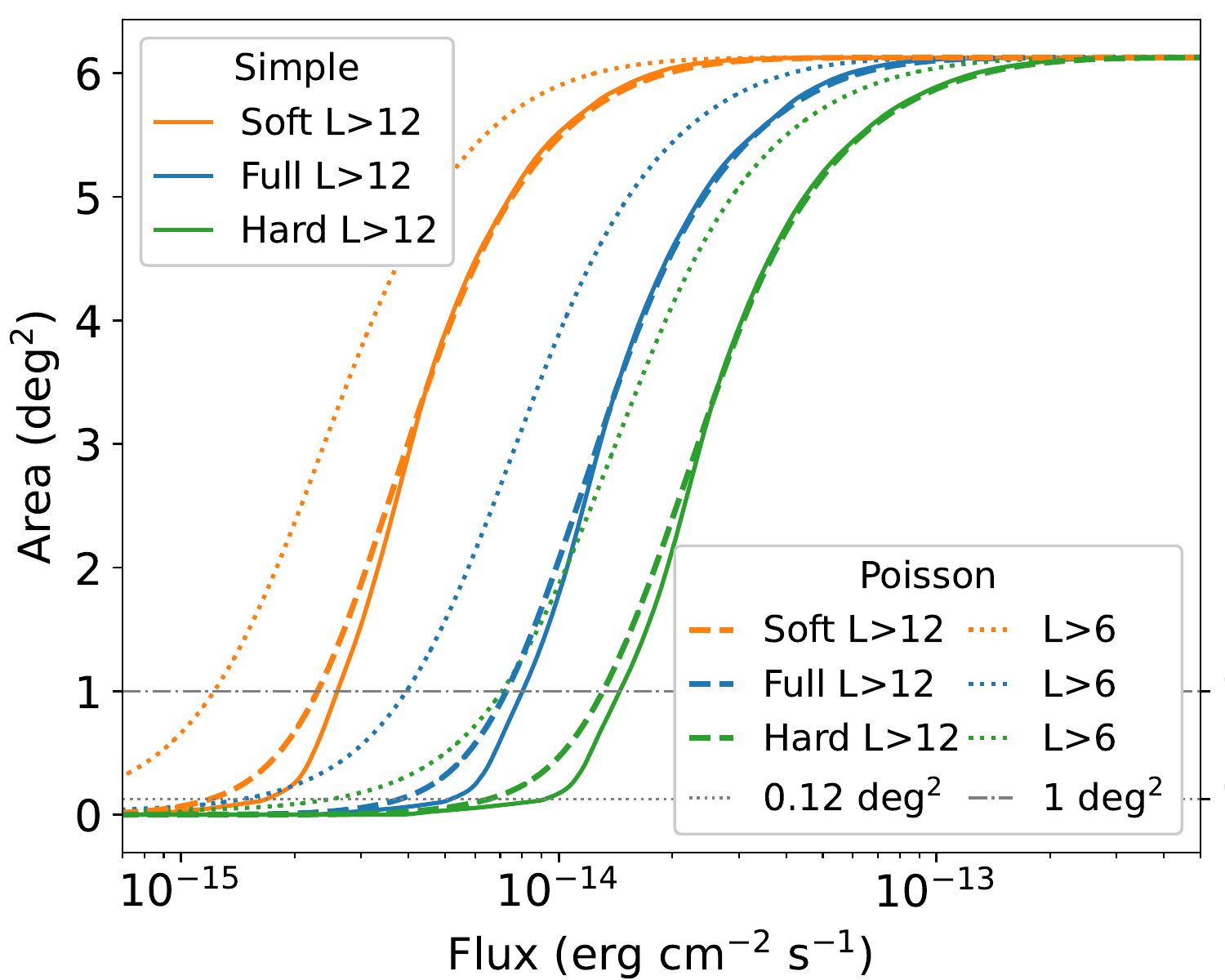}
\caption{
  Sky area coverage in the three bands calculated using the ``Simple'' method adopting a likelihood threshold of 12 (solid lines) and using the ``Poisson'' method adopting likelihood thresholds of 12 (dashed lines) and 6 (dotted lines).
  The grey horizontal lines mark the area of 0.12 and 1 deg$^2$.
}
\label{fig:skycov}
\end{figure}

\begin{figure*}[htb]
\centering
\epsscale{0.9}
\includegraphics[width=0.4\textwidth]{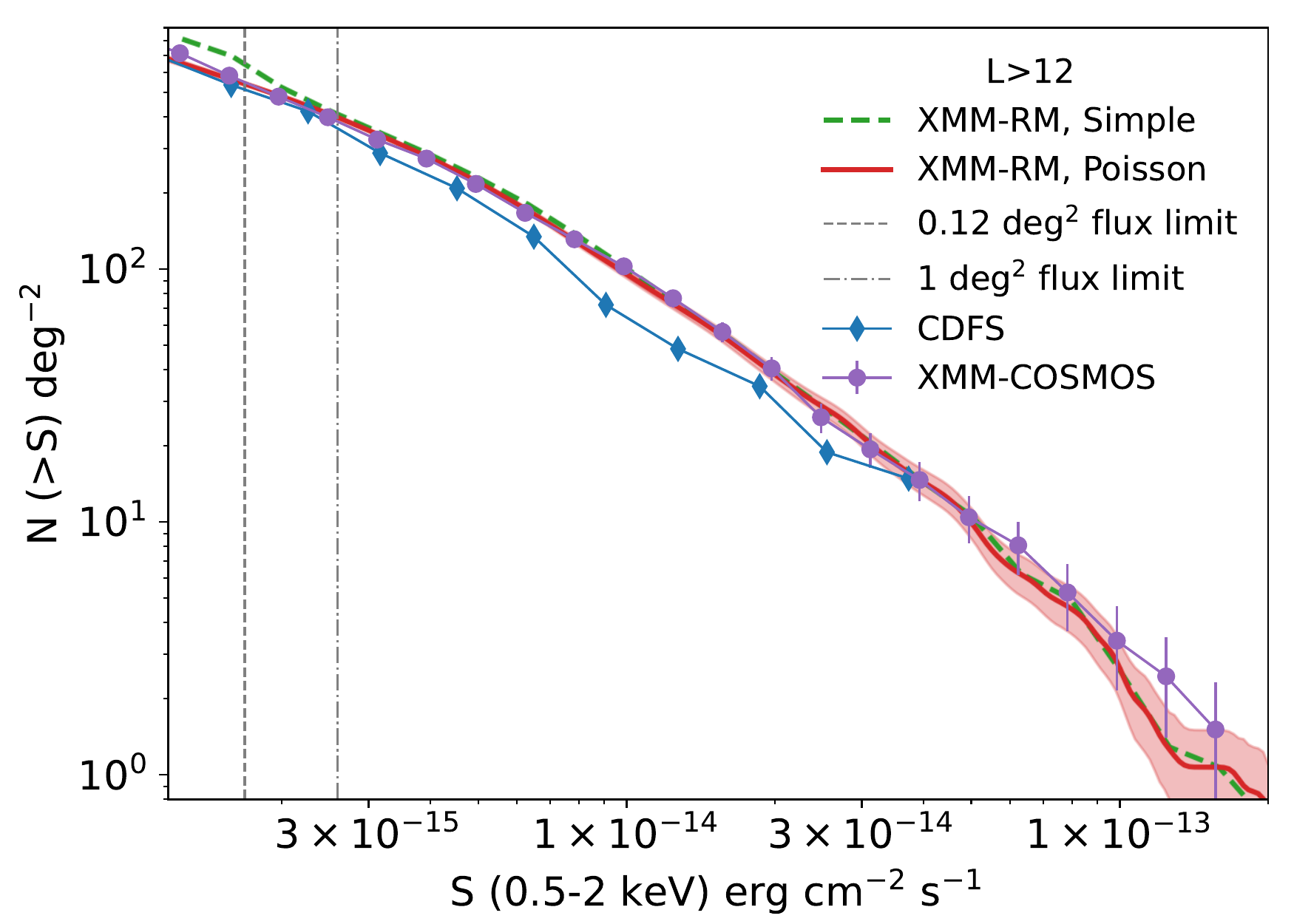}
\includegraphics[width=0.4\textwidth]{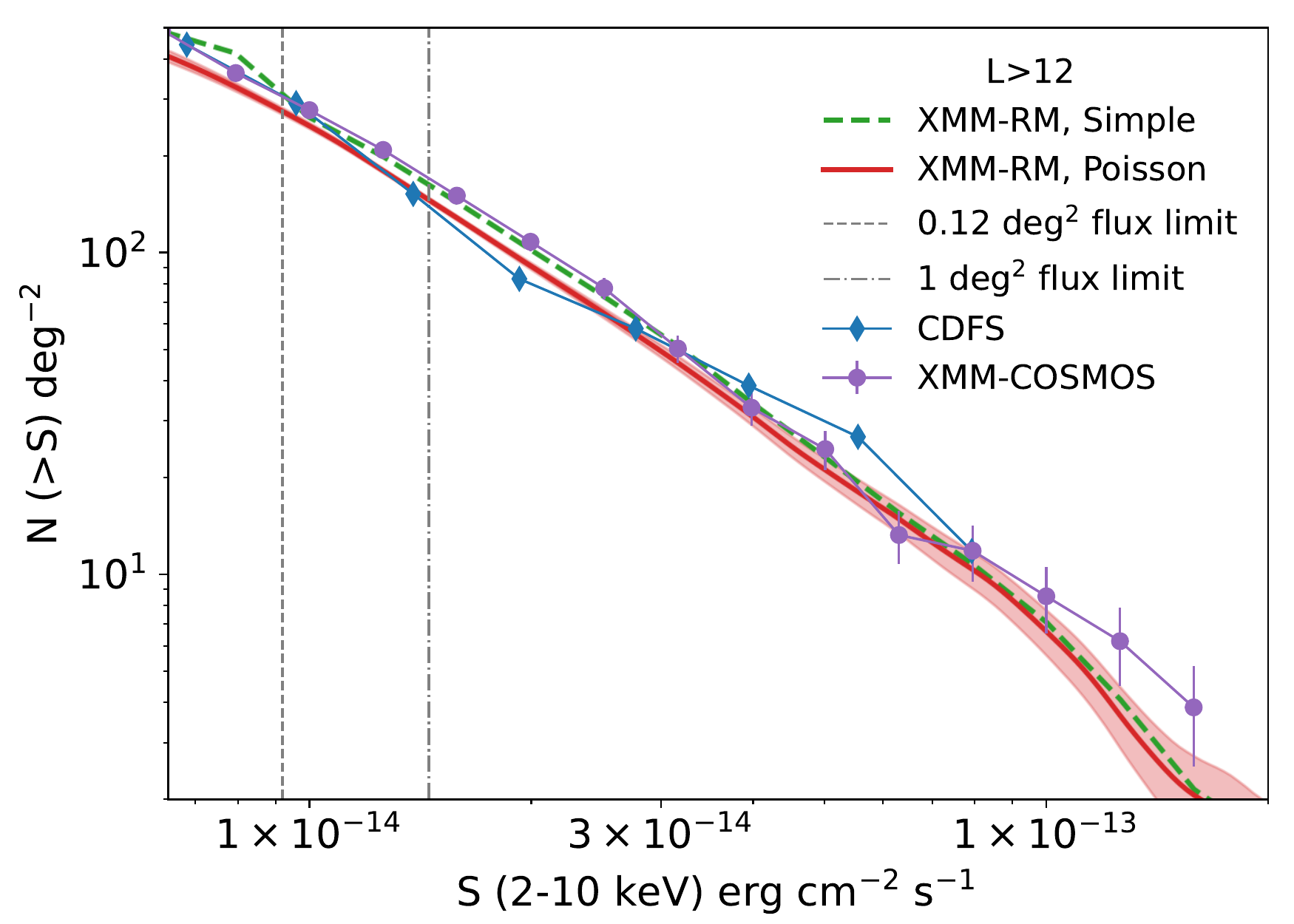}
\includegraphics[width=0.4\textwidth]{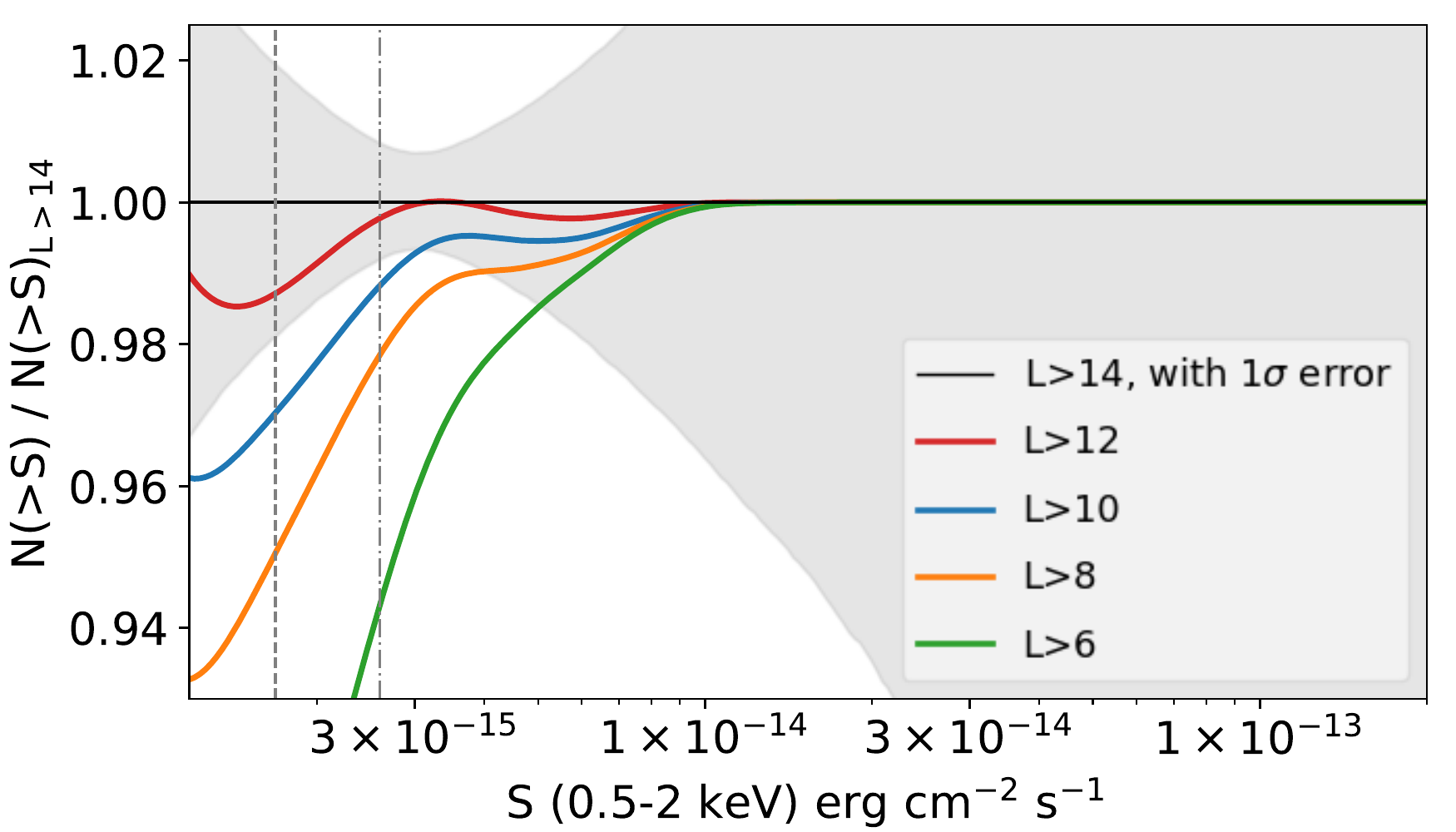}
\includegraphics[width=0.4\textwidth]{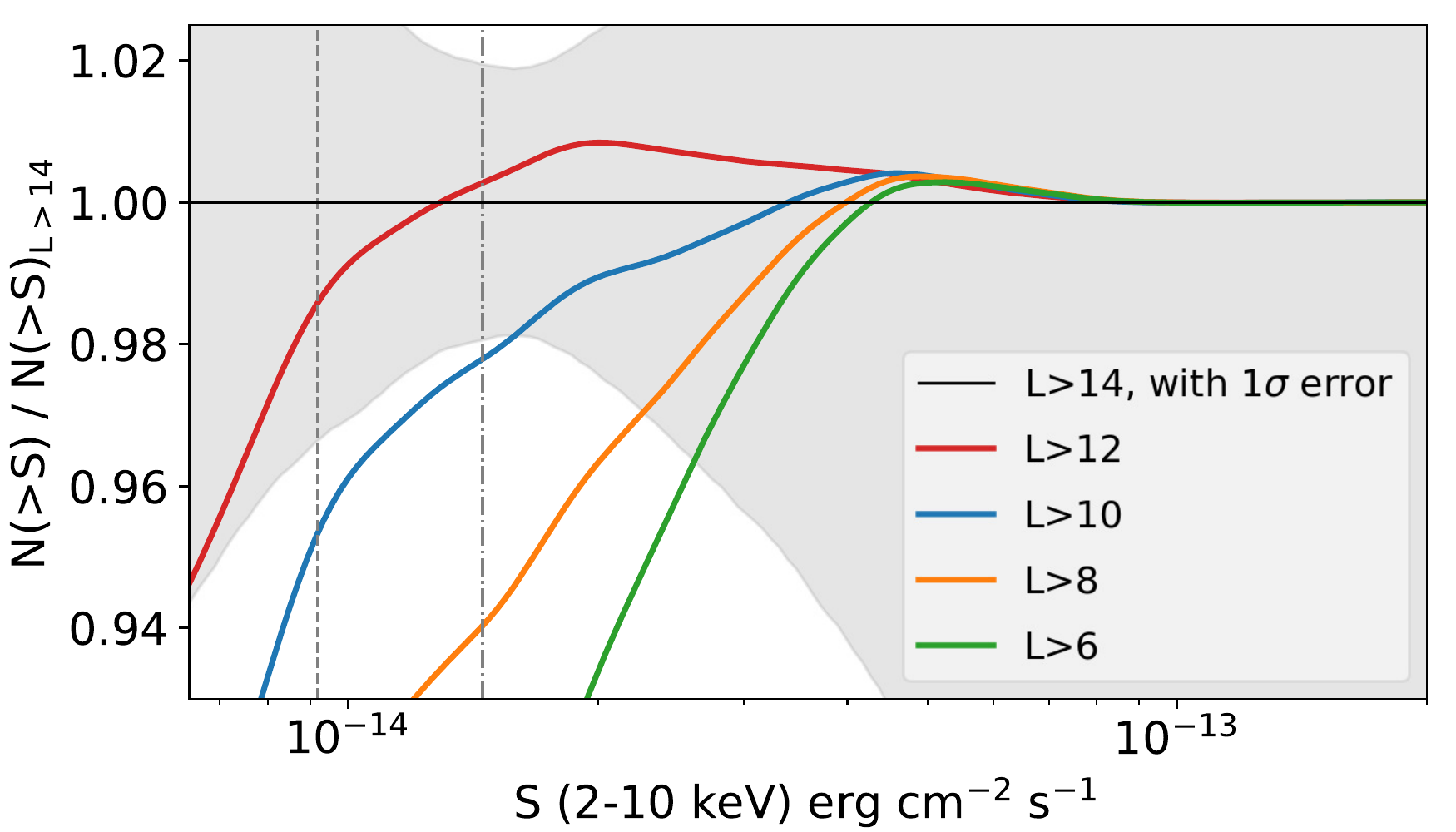}
  \caption{
The upper panels display the soft (left) and hard (right) band logN-logS calculated adopting a likelihood threshold $L>12$ in comparison with that of CDFS \citep[blue diamonds,][]{Luo2017} and XMM-COSMOS \citep[purple points,][]{Cappelluti2009}. The green dashed line is generated using the ``Simple'' number counting method, and the red solid line via the ``Poisson'' probability distribution stacking method. The shaded region displays the $68\%$ confidence interval calculated using Bootstrap percentile method.
The lower panels compare the ``Poisson'' logN-logS calculated adopting $L>12$ (red), $L>10$ (blue), $L>8$ (orange), and $L>6$ (green) with that adopting $L>14$ (black).
In all the panels, the flux limits corresponding to sky areas of 0.12 deg$^2$ (XMM FOV) and 1 deg$^2$ adopting $L>12$ are marked with grey vertical lines.
}
\label{fig:logNlogS}
\end{figure*}

We employed two methods to calculate the number counts of the sources.
The first one is a commonly used method \citep[e.g.,][]{Cappelluti2009} which is called the ``Simple'' number counting method in this work.
It simply sums the number of pixels that reach a given flux limit in the sensitivity map to calculate the sky coverage ($\Omega$) as a function of flux, and then sums the number of sources above a given likelihood threshold, with each source weighted by the reciprocal of the sky coverage at its flux ($f$), i.e.,
\begin{equation}
  \label{equation1}
N(>f) = \sum{\frac{1}{\Omega_{f}}}
\end{equation}
The ``Simple'' sky coverage is shown in Fig.~\ref{fig:skycov} as solid lines, which are used to calculate sample flux limits.
Adopting a Poissonian likelihood threshold of $12$, the flux limits corresponding to a sky area of a circle of radius 12$\arcmin$ (0.12 deg$^2$, approximately the size of the XMM FOV) are $1.7\times 10^{-15}$ and  $9.2\times 10^{-15}$ erg cm$^{-2}$ s$^{-1}$ for the soft and hard band, respectively; and the values corresponding to an area of 1 deg$^2$ are $2.6\times 10^{-15}$ and  $1.45\times 10^{-14}$ erg cm$^{-2}$ s$^{-1}$, respectively.

  The second method -- ``Poisson'' probability distribution stacking -- is more accurate than the ``Simple'' method as it takes into account the Poisson distribution of observed photon counts \citep{Georgakakis2008}.
With a background aperture counts $B$, let $\lambda$ be the expected total aperture counts at a given flux $\lambda = f\cdot Exp \cdot ECF \cdot EEF+B$. The conversion factors from fluxes to counts ($Exp \cdot ECF \cdot ECF$) are obtained from the corresponding maps in Fig.~\ref{fig:EEF_ECF}. 
When converting the detection limit of aperture-source-counts $M$ in each pixel of the sensitivity map into flux, it is converted to a flux probability distribution $Poisson_{\lambda}(\geqslant M+B)$ rather than a single flux limit. The flux distributions in all the pixels are summed to calculate the sky coverage.
As shown in Fig.~\ref{fig:skycov}, such sky coverage curves (dashed) extend to lower fluxes than that using the ``Simple'' method (solid), since with this ``Poisson'' method, any source below the flux limit is considered as detectable with a certain probability.
Likewise, when converting a source aperture counts $S$ to flux, it is converted to a flux probability distribution $Poisson_\lambda(S+B) N(f)$ rather than a single flux.
The factor $N(f)$ is an empirical differential number counts, which is applied in order to correct the Eddington bias.
In this work, we adopt the best-fit broken power-law model of AGN number counts from the 7Ms CDFS survey \citep{Luo2017}.
When summing the sources to calculate the logN-logS as in Equation~\ref{equation1},
the number 1 of each source is replaced with the normalized flux probability distribution $P(f)df/\int{P(f)df}$.

  It is common practice to use a clean sample with a high selection threshold to calculate the logN-logS.
To limit the contamination in the sample to 1\%, a common choice of Poissonian likelihood threshold is 12.43 ($-\ln 4\times 10^{-6}$) \citep{Georgakakis2008,Laird2009,Georgakakis2011}.
To understand the affection of the threshold choice, we test a few thresholds of $6$, $8$, $10$, $12$, and $14$.
Across the XMM-RM field, the exposure depth varies by orders of magnitude. The very-shallow part of the field has few photons and thus huge uncertainties.
To calculate the logN-logS, we mask out such regions by applying a lower limit of $4\times 10^{15}$ on the soft-band Exp-ECF map, which excludes $14\%$ of the sources and $24\%$ of the area.
We also exclude sources classified as extended or attributed to galaxy clusters (see \S~\ref{sec:clusters}).
Adopting a likelihood threshold of $12$, the soft and hard band sub-samples include $1557$ and $564$ sources, respectively.

  Fig.~\ref{fig:logNlogS} displays the cumulative number counts calculated adopting $L>12$ in comparison with that from the 7Ms CDFS \citep{Luo2017} and XMM-COSMOS \citep{Cappelluti2009}, which are corrected for the different energy bands and spectral models used.
  The ``Simple'' number counts are broadly consistent with that of XMM-COSMOS within uncertainties. The CDFS counts show some deviations, which are most likely explained by sampling variance, given the very small area of the CDFS field. 
  Having the Eddington bias corrected, the number counts measured with the ``Poisson'' method is slightly lower than that with the ``Simple'' method. This correction is more significant in the hard band, which has a higher background and thus larger relative uncertainties.

The number counts calculated adopting different thresholds are compared in the lower panels of Fig.~\ref{fig:logNlogS}.
Reducing the threshold, the sample is expected to contain more contamination from either spurious sources (fluctuations) or extended sources.
However, on the contrary, the number counts become lower with lower thresholds, especially in the hard band where the relative uncertainties are larger.
As discussed in \S~\ref{sec:poissonianlikelihood}, because of the large scatter of likelihoods, the PSF-fitting likelihood selected sample is incomplete with respect to the Poissonian likelihood. Although we have minimized such incompleteness by choosing a low PSF-fitting likelihood threshold, at low likelihoods, the ``Poisson'' method aggravates such incompleteness -- a fraction of below-flux-limit sources are considered as detectable through Poisson fluctuation but actually cannot survive the PSF-fitting pre-selection.
Therefore, a high threshold is required to recover the logN-logS not only in order to avoid spurious sources.
With the ``Poisson'' method, more prominent problems to avoid are the sample incompleteness and the large uncertainties at low thresholds.
Above a flux limit that corresponds to an area of 0.12 deg$^2$, the $L>12$ and $L>14$ cases are highly identical with only a $\sim 1\%$ difference. With $L>10$, the difference of logN-logS from that of $L>14$ starts to be larger than the uncertainty. Therefore, we suggest to adopt a threshold of 12.

\section{Multi-band Counterparts}
\label{sec:ctp}

\subsection{Combined Legacy--unWISE catalog}
We identify multi-band counterparts of the XMM-RM sources from the optical Legacy \citep{Dey2019} and the IR unWISE \citep{Schlafly2019} catalogs.
The Legacy catalog presents $g$, $r$, and $z$ band magnitudes for each source.
The unWISE catalog provides the $W1$ (3.4$\micron$) and $W2$ (4.6$\micron$) magnitudes, although some sources are detected in only one of the two bands.
Considering that the typical XMM positional uncertainty ($\sim 2\arcsec$, half pixel size) is much larger than that of the optical/IR positions, we first combine the Legacy and the unWISE catalogs and then match the XMM sources to the combined catalog.
In order to take account of the unknown systematic uncertainty of the Legacy catalog and to avoid extremely small and potentially unrealistic positional uncertainties, we add an additional systematic positional uncertainty of 0.3$\arcsec$ in quadrature to both the Legacy and the unWISE sources.
This may overestimate the positional errors of some optical sources, but it prevents failures in identifying optical-IR match of sources caused by underestimation of positional errors in some cases, and it is small enough not to cause any problem in the X-ray counterpart association.
As the optical positional accuracy is the best, we ran NWAY to associate the unWISE sources to the Legacy ones within 4$\arcsec$, using the Legacy $z$ band magnitude prior generated through the NWAY ``AUTO'' method.
We select the matches with $p_{any}>0.02$, which corresponds to a $5\%$ false association rate according to a Monte Carlo test.
Based on these matches, we merged all the Legacy and unWISE sources into one catalog, including 41\% Legacy--unWISE pairs, 44\% Legacy-only sources, and 15\% unWISE-only sources.
We adopt an upper limit of 24 on the IR $W1$ and $W2$ magnitudes and an upper limit of 29 on the optical $g$, $r$, $z$ magnitudes.
The magnitudes of undetected sources are set to these upper limits.
As discussed later, the simple Monte Carlo method we used leads to an overestimated false rate which is boosted high by the high density of optical sources.
Because the optical-IR positional uncertainty is much better than that of X-ray, potential false matches and mis-matches caused by abnormal positional accuracy in some rare cases cannot cause any visible problem in this work -- anyway, all the optical-IR sources are available in the stacked catalog to be picked as counterparts of X-ray sources.

\subsection{Priors for X-ray sources}
\label{sec:colorprior}

In order to identify optical/IR counterparts of the XMM-RM sources efficiently using the Bayesian method NWAY \citep{Salvato2018}, we create optical/IR color and magnitude priors on the basis of the Chandra catalogs of C-COSMOS \citep{Marchesi2016} and Stripe-82 \citep{LaMassa2016}. Chandra sources have excellent positional accuracy ($\sim 0.5\arcsec$) and can easily be matched to the correct counterparts.
We select the unWISE sources within 30$\arcsec$ of the Chandra sources and the Legacy sources within 35$\arcsec$ and match them as done above for the XMM-RM field. 
We also select the counterparts with $p_{any}>0.02$, which corresponds to a slightly higher false rate of $6\%$ in the COSMOS field because it has a deeper Legacy coverage than the XMM-RM field.
From the combined optical/IR catalog selected within 30$\arcsec$ of Chandra sources, we search for IR/optical counterparts using NWAY within a maximum distance of 4$\arcsec$ using the Legacy $g$ band and the unWISE $W1$ band priors generated through the NWAY ``AUTO'' method.
We select only the reliable counterparts with $p_{any}>0.6$ (corresponding to a false rate of $5\%$) and $p_{i}>0.6$.
In order to have similar fluxes to our sample, we select only sources with soft (0.5-2 keV) fluxes $>1.0\times 10^{-15}$ erg\ cm$^{-2}$\ s$^{-1}$; see Fig.~\ref{fig:softflux} for the soft flux distributions of the XMM-RM sources and the C-COSMOS and Stripe-82 Chandra sources.
The selected optical/IR counterparts are compared with the other sources in the parent sample to build priors as follows.

\begin{figure}[htbp]
\epsscale{1}
\plotone{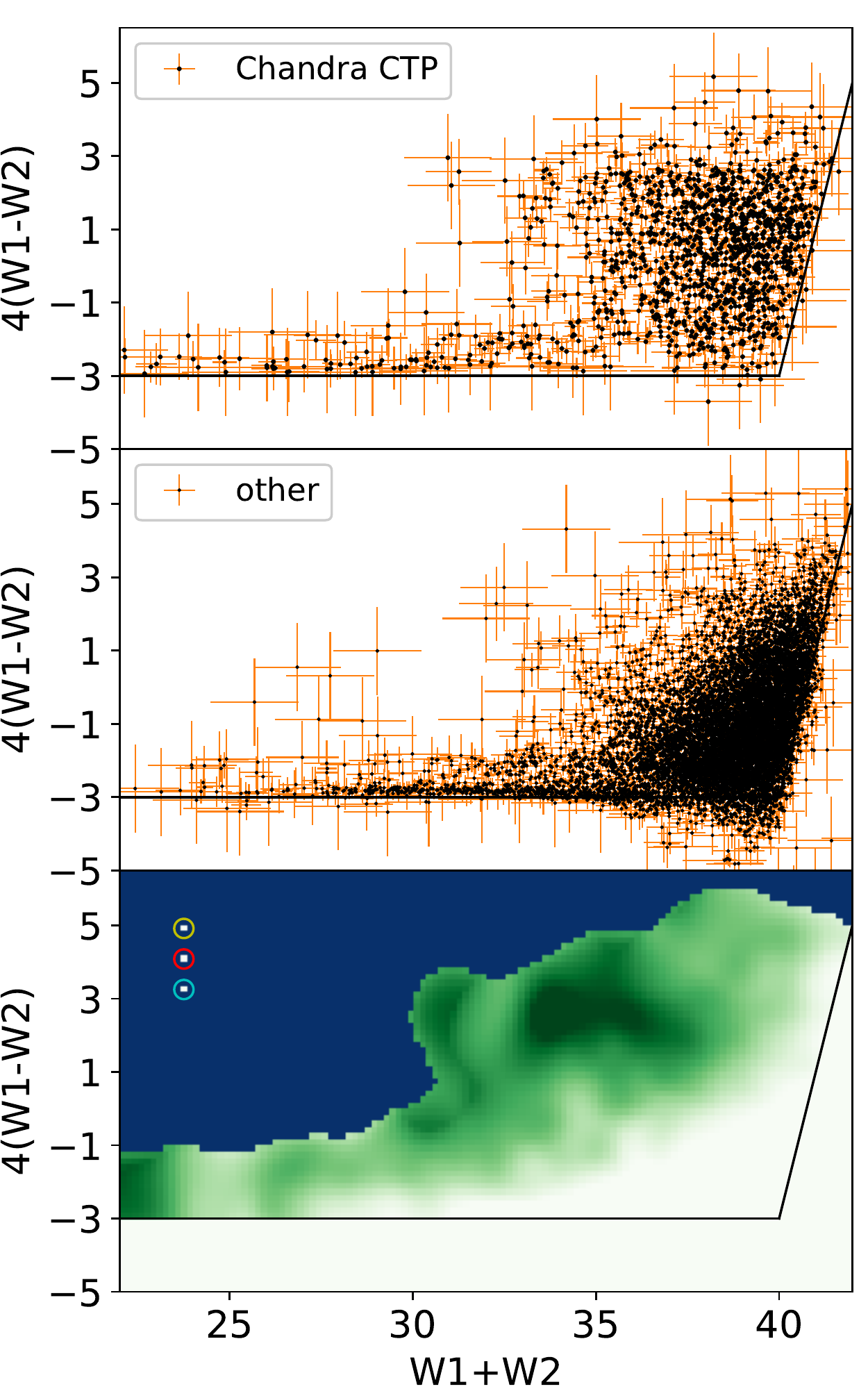}
\caption{The top and middle panels display the distributions of the unWISE counterparts of Chandra sources and the other unWISE sources within 30$\arcsec$ of Chandra sources. The errorbar corresponds to the square root of the Voronoi cell area of each point.
  The bottom panel presents the color-coded prior in terms of the ratio between the smoothed distributions of the top and middle panels, from low to high value in green color from light to dark. The blue region is not used. The three special pixels marked with red, yellow, and cyan circles correspond to unWISE sources detected only in W1, only in W2, and unWISE undetected sources, respectively.
The two axes have the same scaling from data to plotting units.}
\label{fig:unWISEprior}
\end{figure}

We define an IR color in the space of $W1+W2$ $\sim$ $4(W1-W2)$, where $W1,W2$ are the unWISE magnitudes and the factor $4$ is added to stretch the distribution in the $W1-W2$ direction just in order to preserve the $W1-W2$ gradient during smoothing.
Fig.~\ref{fig:unWISEprior} displays the colors of the selected counterparts (top panel) and the other sources within 30$\arcsec$ of the Chandra source positions (middle panel).
Using the Python package ``SweeplineVT'' \citep{Liu2013}, Voronoi tessellation is run on these points and the Voronoi cell area of each point is calculated. The square root of the cell area of each point is indicated as error bars in Fig.~\ref{fig:unWISEprior}, with a maximum cut of 1.2 applied.
We pixelate the space in the ranges of $22<W1+W2<44$ and $-7.5<4(W1-W2)<6.5$ into $132\times84$ pixels, which guarantees a high spatial resolution of the 2D source distribution.
For each point, we fill into the pixelated image a 2D Gaussian probability density function with a scale of the square root of its cell area (error bars in Fig.~\ref{fig:unWISEprior}).
Limited by the sample size, the distribution is not sufficiently smoothed across the space. Thus we adopt a further Gaussian smoothing using a scale of 2 pixels.
The ratio between the smoothed distributions (bottom panel of Fig.~\ref{fig:unWISEprior}) is used as the prior for counterparts of X-ray sources.

As shown in Fig.~\ref{fig:unWISEprior}, we manually draw a horizontal line at $4(W1-W2)=-3$ and a line with a slope of 4 (black lines), and consider the region below these lines as a forbidden region by setting the probability of this region to the minimum value in the prior distribution.
We also add three special pixels in the prior image, which are marked with red, yellow and cyan circles in Fig.~\ref{fig:unWISEprior}, in order to store the values for three special cases: the unWISE sources detected only in the W1 band (red) and only in the W2 band (yellow) and the unWISE undetected sources (cyan).
Taking these three special categories into account, this magnitude-color prior covers almost the entire parameter space. The uncovered region (blue region in Fig.~\ref{fig:unWISEprior}) contains barely any sources.

\begin{figure}[htbp]
\epsscale{0.9}
\plotone{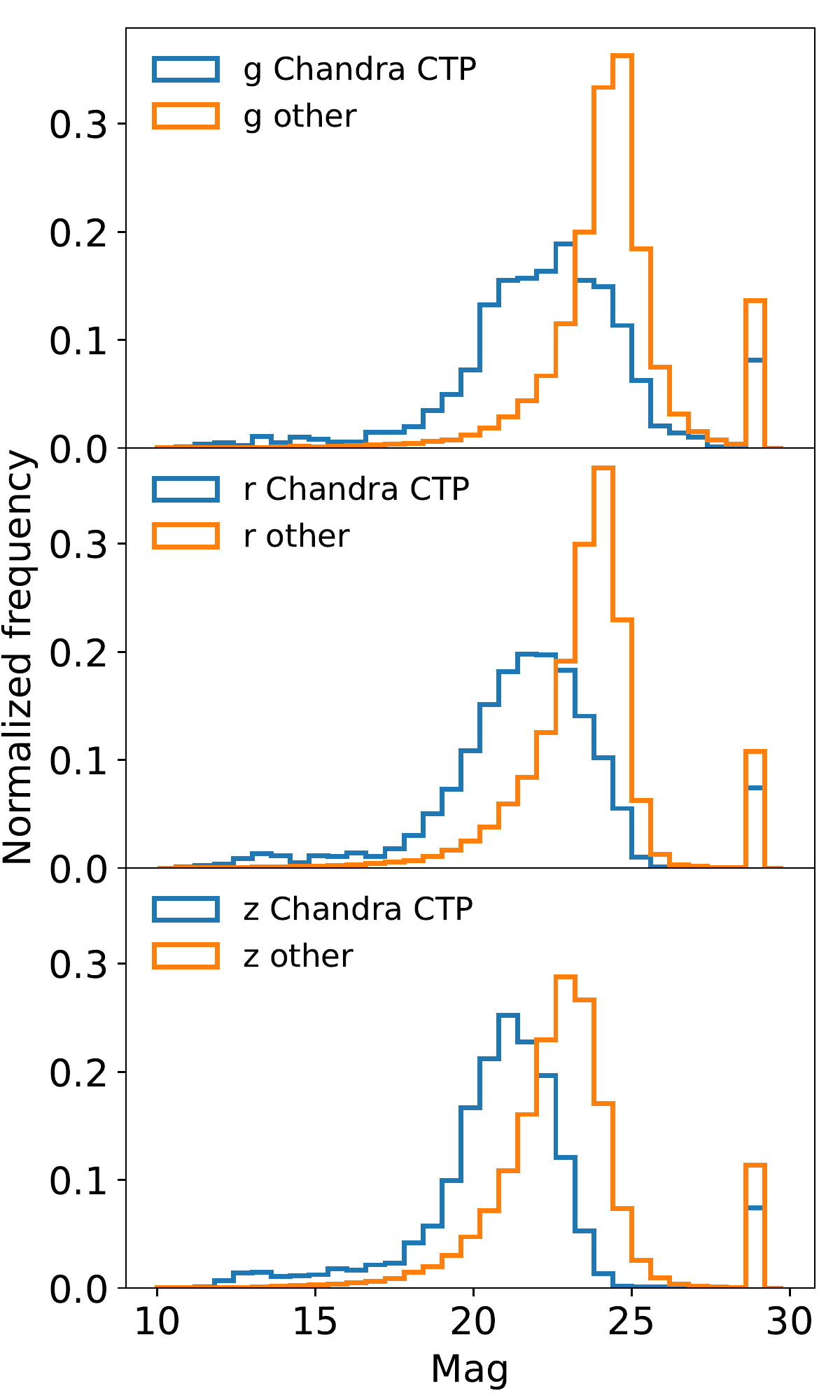}
\caption{The normalized $g$, $r$, $z$ band magnitude distributions of the reliable counterparts of Chandra sources (blue) and the other Legacy sources (orange).
}
\label{fig:Legacy_mag_priors}
\end{figure}

Fig.~\ref{fig:Legacy_mag_priors} displays the normalized Legacy $g$, $r$, $z$ band magnitude distributions of the reliable counterparts of Chandra sources (blue) and the other Legacy sources (orange).
These magnitude distributions can also be used as priors for selecting counterparts of X-ray sources.
We do not create an optical color prior as done above for the unWISE color, because it is likely to introduce a bias against type II AGN.

\subsection{Optical/IR counterparts of XMM-RM sources}
\label{sec:CTP}

\begin{figure*}[htbp]
\epsscale{0.8}
\plotone{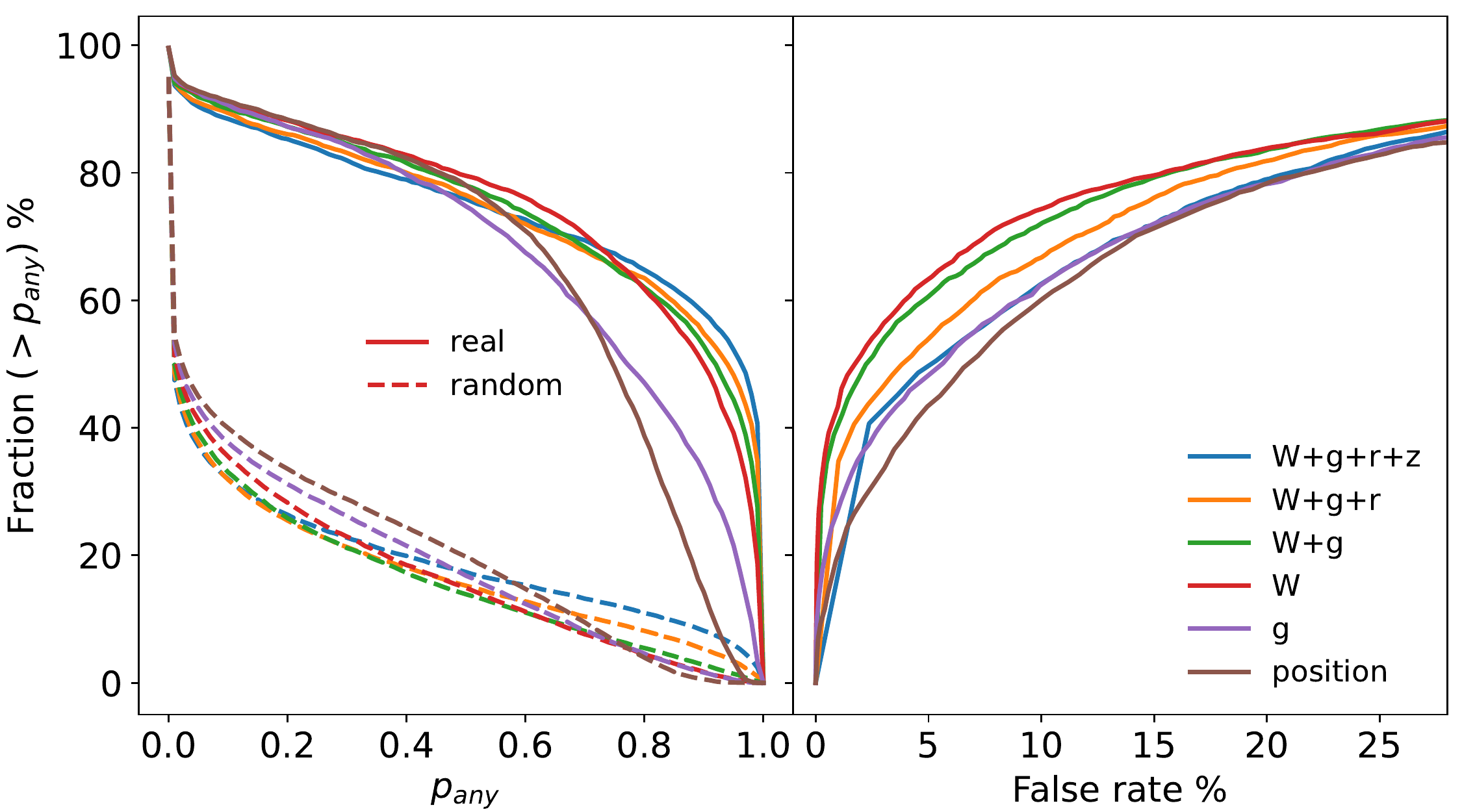}
\caption{The left panel displays the cumulative distributions of $p_{any}$ of real sources (solid lines) and of random positions (dashed lines), which represent the completeness and false rate of the counterpart selection. The right panel shows the completeness as a function of false rate with a varying $p_{any}$ threshold.
The cases of adopted priors include: no additional prior (position only), $g$ band magnitude prior ($g$), unWISE prior ($W$), and unWISE prior plus one or multiple magnitude priors ($W+g$, $W+g+r$, $W+g+r+z$).
}
\label{fig:nway_results_comp}
\end{figure*}

Having established the prior based on the Chandra data, we turn back to the XMM-RM field. From the Legacy--unWISE combined catalog, 
we select the sources within 30$\arcsec$ of all the XMM sources as the candidate sample, whose total area is 0.74 deg$^2$.
We run NWAY with a completeness prior of $0.7$ and a maximum distance of 15$\arcsec$, using a series of prior choices as described below.
As a Monte Carlo test, we re-distribute the XMM-RM sources randomly in the XMM-RM footprint.
  We maintain a minimum separation of 10$\arcsec$ among the random positions, but do not purposely keep away from the position of real sources.
  The test is repeated five times to improve the statistics.
The completeness and false rate at any $p_{any}$ threshold based on the different prior choices are compared in Fig.~\ref{fig:nway_results_comp}. Achieving a higher completeness and a lower false rate at the same time indicates a more efficient identification.
First, we run without any additional prior, e.g., based on only positions and their uncertainties.
Then we use the $g$ band prior created above using Chandra sources. This improves the identification efficiency slightly.
Replacing the $g$ prior with the unWISE prior ($W$), we find a significant improvement.
Adopting the $W$ prior, we then tried adding 1) the $g$ prior, 2) the $g$ and $r$ priors, and 3) the $g$, $r$, and $z$ priors.
We find that the more optical priors we add, the less efficient the identification is.
This is because the magnitudes in different bands are correlated, so adding them does not introduce much additional independent information. On the other hand, adding further unnecessary priors pushes up the $p_{any}$ values of bright sources which are occasionally matched to random positions.
Therefore, we present the NWAY results using only the unWISE prior.

One X-ray source can have multiple SDSS/unWISE counterparts identified by NWAY.
For the sake of completeness, all the counterparts from NWAY using the unWISE prior are provided along with this paper (see Appendix~\ref{app:Xraycolumns}).

\begin{figure}[htbp]
\epsscale{1.15}
\plotone{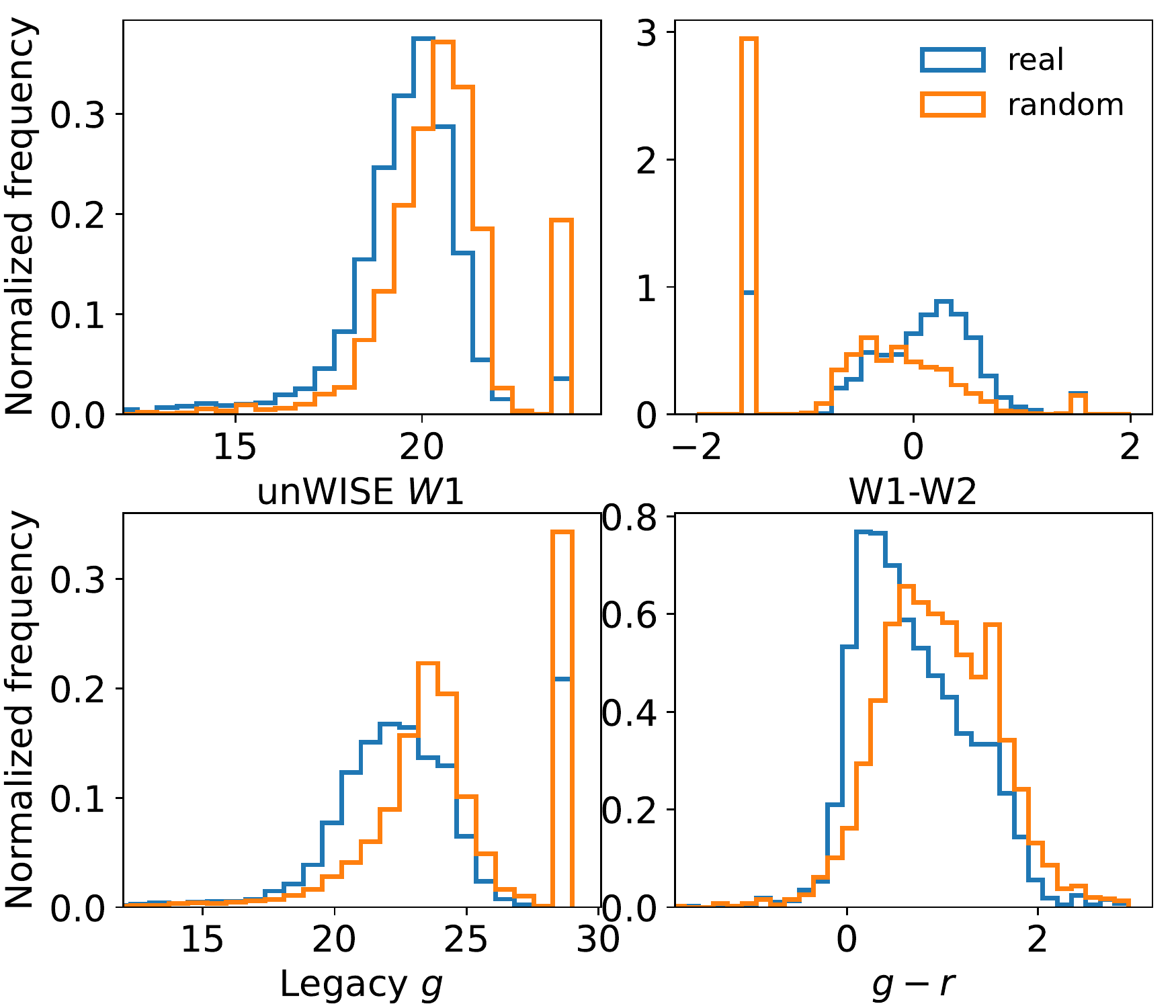}
\caption{Comparison of magnitudes (unWISE W1, Legacy $g$) and colors (W1-W2,$g-r$) of the best counterparts of the real sources (blue) and random positions (orange) with $p_{any}>0.36$.
  Legacy-undetected sources are set as $g=29$; unWISE-undetected sources are set as $W1=24$.
  unWISE sources detected in only one of the W1 and W2 bands are plotted as $W1-W2=1.5$ or $-1.5$.}
\label{fig:diff_real_random}
\end{figure}

As shown in Fig.~\ref{fig:nway_results_comp}, selecting the best NWAY counterparts with $p_{any}$ above 0.78, 0.63, and 0.36 lead to a completeness of 63\%, 74\%, and 84\% with corresponding false counterpart identification rates of 5\%, 10\%, and 20\%, respectively.
Here the false rate is defined as the probability of finding any optical/IR counterpart at a random position.
A high false rate measured by the Monte Carlo test is partially caused by the large XMM positional uncertainty.
More importantly, it is because the Legacy and unWISE surveys are both very deep and combining them results in a high density of sources.
We remark that such a false rate is much higher than the genuine probability of one counterpart being false, because in the highly-crowed optical/IR catalog, many sources are not potential counterparts of X-ray sources.
As shown in Fig.~\ref{fig:diff_real_random}, adopting $p_{any}>0.36$, the best counterparts of real sources and of random positions are significantly different in magnitudes and colors. The difference is large even if adopting a strict threshold of $p_{any}>0.78$.
The counterparts of random positions tend to be fainter and more likely undetectable in some bands.
Therefore, we recommend to allow a relatively higher false rate for these counterparts, e.g., 10\% or 20\%, rather than 5\% or 1\%.

\subsection{Astrometric accuracy}
\label{sec:astrometricaccuracy}
\begin{figure}[htbp]
\epsscale{1}
\plotone{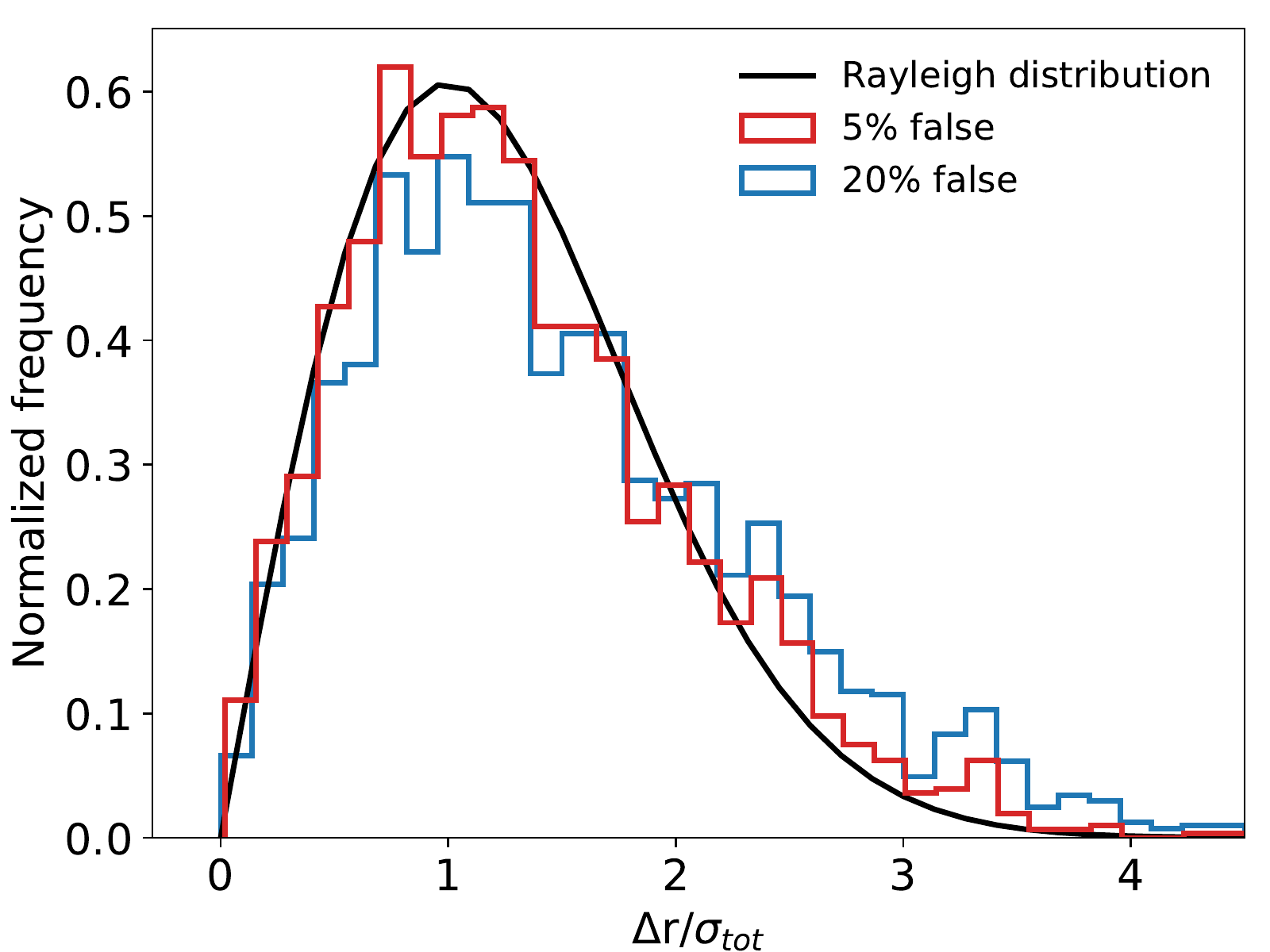}
\caption{The distributions of X-ray to optical/IR position separation $\Delta r$ in units of total 1-dimensional positional uncertainty $\sigma_{tot}=\sqrt{\sigma^2_X/2+\sigma^2_{opt/IR}/2}$.
The red and blue lines indicate selections of $p_{any}>0.78$ and $p_{any}>0.36$, respectively. The black line is the Rayleigh distribution.
}
\label{fig:Rayleigh}
\end{figure}

To test if an additional systematic positional uncertainty is present, we compare the distribution of X-ray -- optical/IR source separation with the Rayleigh distribution in Fig.~\ref{fig:Rayleigh}.
This method relies on the fact that the distribution of the ratio $x = \Delta r/\sigma_{tot}$, where $\Delta r$ is the separation between the X-ray source and its optical/IR counterpart and $\sigma_{tot}$ is the total 1-dimensional positional error $\sqrt{\sigma^2_X/2+\sigma^2_{opt/IR}/2}$, should follow the Rayleigh distribution $f(x) = x\ exp(-x^2/2)$, as long as $\sigma_{tot}$ is an accurate estimate of the true positional error \citep{Watson2009,Rosen2016,Pineau2017}.
Adopting a $p_{any}$ threshold that corresponds to a 20\% false rate (blue line), we have an excess tail above the Rayleigh distribution at high $x$.
Adopting a higher threshold that corresponds to a 5\% false rate (red line), the distribution fits the Rayleigh distribution much better.
These results are broadly consistent with that found by \citet{Rosen2016} for the 3XMM catalog; see a more detailed discussion therein.
The tail excess can be caused by spurious matches and/or underestimation of positional uncertainties in some cases probably at high off-axis angles.
For the majority of the catalog, the positional uncertainties are accurate and need no additional correction.

\subsection{SDSS spectra and best counterparts}
\label{sec:SDSSspec}

We search for SDSS spectra of the optical/IR counterparts from the DR16 catalog \citep{York2000} through a positional match within a radius of 1$\arcsec$ (BOSS fiber radius) and find $1206$ SDSS spectra for $1141$ X-ray sources, including $847$ AGN, $312$ galaxies, and $47$ stars.
$71\%$ of the AGN ($603$) are in the SDSS-RM quasar catalog \citep{Shen2019}.
There are five sources with SDSS spectroscopic redshifts $>5$. Since such high redshifts from the SDSS spectroscopic pipeline are often unreliable, we visually examined these spectra. Better fits with smaller redshifts are found for two of them (both $z \approx 2$). We replace their SDSS redshifts with the manually fitted ones. For the other three, the spectra are too noisy to provide robust redshift measurements. We exclude them from further analysis by multiplying their SDSS redshifts by -1.

NWAY calculates $p_i$ on the basis of priors which are generated considering all the X-ray sources as a population with similar magnitude and color distributions, regardless of source types. Such priors are a good choice for the majority of the sample, but not necessarily appropriate for each individual source.
Specifically, the priors favor optical/IR sources which appear brighter. Although AGN are more likely X-ray emitters than stars, stars which are often brighter in optical/IR will be assigned higher probabilities according to the priors.
Therefore, rather than simply selecting the best counterpart as the one with $match\_flag==1$, which means the highest $p_i$ in NWAY, we select the best counterpart for each X-ray source taking into account the SDSS spectra.

In $915$ cases, the SDSS spectroscopic objects are identified as the best ($match\_flag==1$) counterparts of X-ray sources. In the other $226$ cases, a better counterpart with a higher $p_i$ than the SDSS spectroscopic object is found in the unWISE-Legacy catalog, which is deeper than the SDSS catalog.
We note that in $17$ such cases, the position-based posterior matching probability ($dist\_post$) of the SDSS spectroscopically confirmed AGN is even higher than or at least approximately equal to that of the best unWISE-Legacy counterpart, indicating such SDSS AGN have relatively lower $p_i$ than other unWISE-Legacy sources only because of the priors adopted.
However, it should be considered as a strong additional prior that AGN tend to be X-ray emitters.
Therefore, we select the SDSS objects as the best counterparts of these $17$ sources; and in other cases, we select the one with $match\_flag==1$.

Among the selected best counterparts, there are $932$ SDSS spectroscopic objects. For the ones in SDSS-RM, we adopt the redshift and class (AGN) from SDSS-RM \citep{Shen2019}, which were examined carefully, instead of the pipeline results of the SDSS-DR16 catalog.
We also visually inspect the non-AGN SDSS spectra and find $21$ sources have both a high X-ray luminosity ($\log L_X>42.5$ erg s$^{-1}$) and a type-II-AGN-like spectrum (e.g., with a strong narrow [OIII] emission line). Their classes are manually changed to AGN.
Eventually, there are $831$ AGN ($594$ SDSS-RM quasars), $96$ galaxies, and $5$ stars.

\begin{figure}[htbp]
\epsscale{1}
\plotone{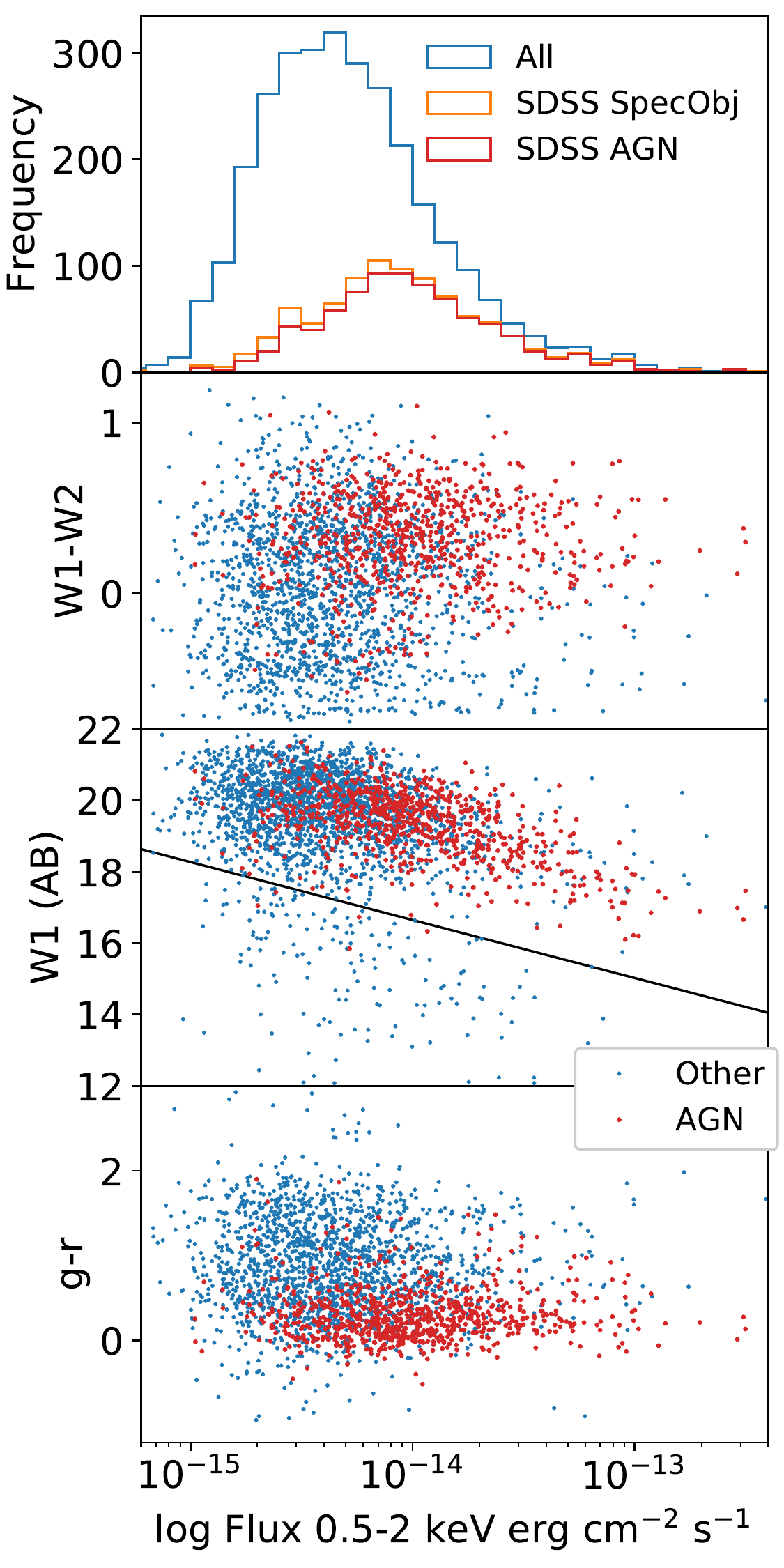}
\caption{
  The top panel displays the X-ray flux distributions of all the sources (blue), the SDSS spectroscopically observed ones (orange), and the spectroscopically confirmed AGN (red).
  The other panels display the unWISE $W1-W2$ color (top), the $W1$ AB magnitude (middle), and the SDSS $g-r$ color (bottom) of the best counterparts as a function of the X-ray fluxes. A source is plotted in red if classified as AGN.
  The black line is the empirical line from \citet{Salvato2018} to separate AGN from galaxies and stars.
}
\label{fig:ctp_color}
\end{figure}

Fig.~\ref{fig:ctp_color} displays the unWISE $W1-W2$ color, the $W1$ AB magnitude, and the SDSS $g-r$ color of the best counterparts as a function of the soft band X-ray fluxes. The spectroscopically confirmed AGN have relatively bluer optical color and redder IR color, and almost all of them lie above the empirical line suggested by \citet{Salvato2018} to separate AGN from normal galaxies and stars ($W1_{\mathrm{Vega}}=-1.625\times \log$Flux$_{\mathrm{0.5-2keV}}-8.8$).

\begin{figure}[htbp]
\epsscale{1}
\plotone{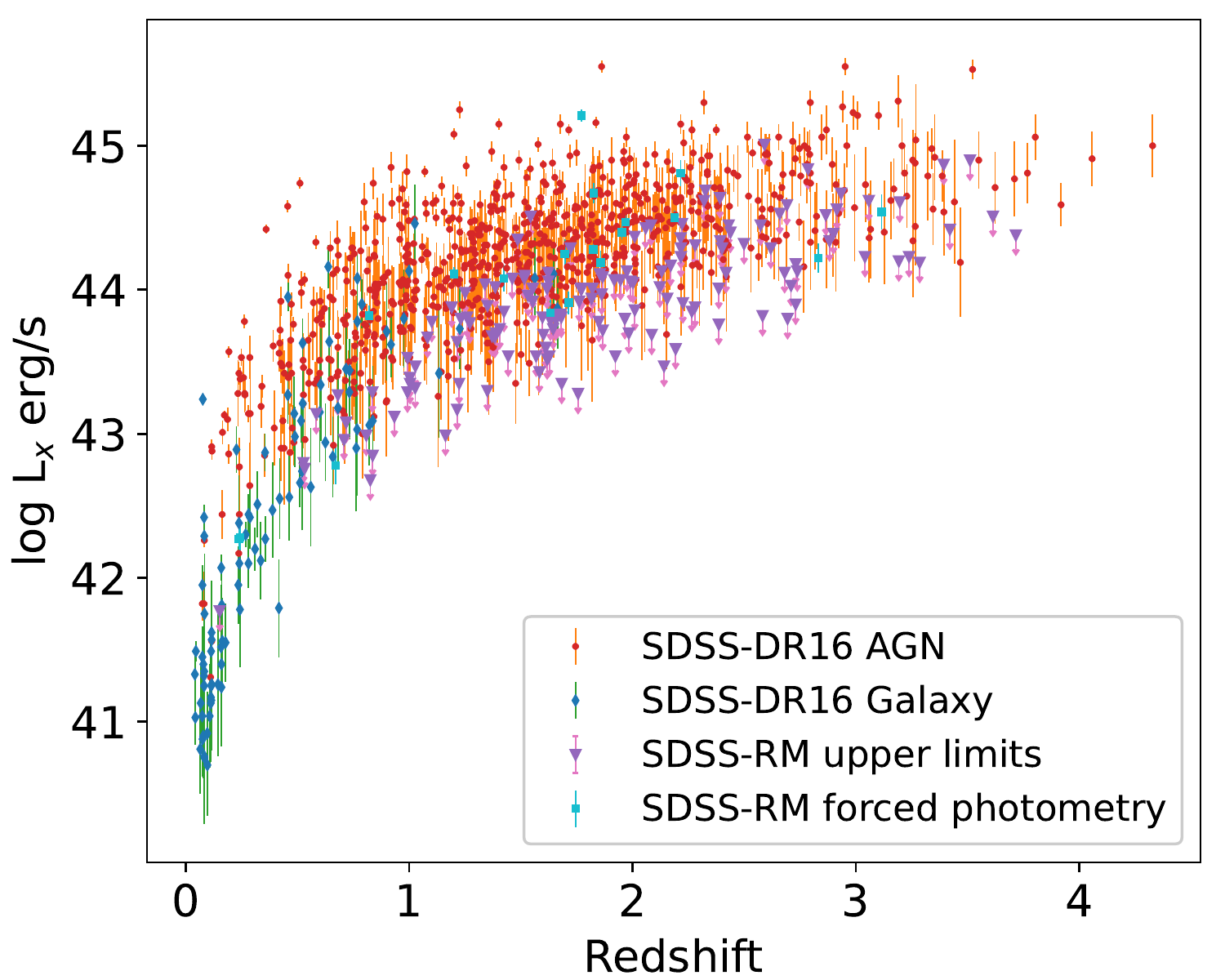}
\caption{
  The distribution of the rest-frame 2-10 keV luminosity and SDSS spectroscopic redshifts.
  The red circular points indicate spectroscopically confirmed AGN; the blue diamonds indicate galaxies; the cyan squares are measured through forced photometry at the positions of SDSS-RM quasars; and the purple triangles are upper limits of X-ray undetected SDSS-RM quasars.
}
\label{fig:L_z}
\end{figure}

For the AGN and galaxies with SDSS spectroscopic redshifts, we calculate the rest-frame 2-10 keV luminosities assuming a power-law with a photon index of 1.7 and Galactic absorption \citep{HI4PI}. In general, the soft band flux is preferentially used to calculate the luminosity. However, at $z<1$, we choose the hard band to avoid a large k-correction as long as the hard band relative flux uncertainty (ratio of flux error to flux value)  $\Delta f_{Hard}$ is lower than $0.7$ or $\Delta f_{Soft}+0.2$, where $\Delta f_{Soft}$ is that in the soft band.
The luminosity -- redshift distribution of these sources is shown in Fig.~\ref{fig:L_z}.

\subsection{Undetected SDSS-RM quasars}
\begin{figure}[htbp]
\epsscale{1.15}
\plotone{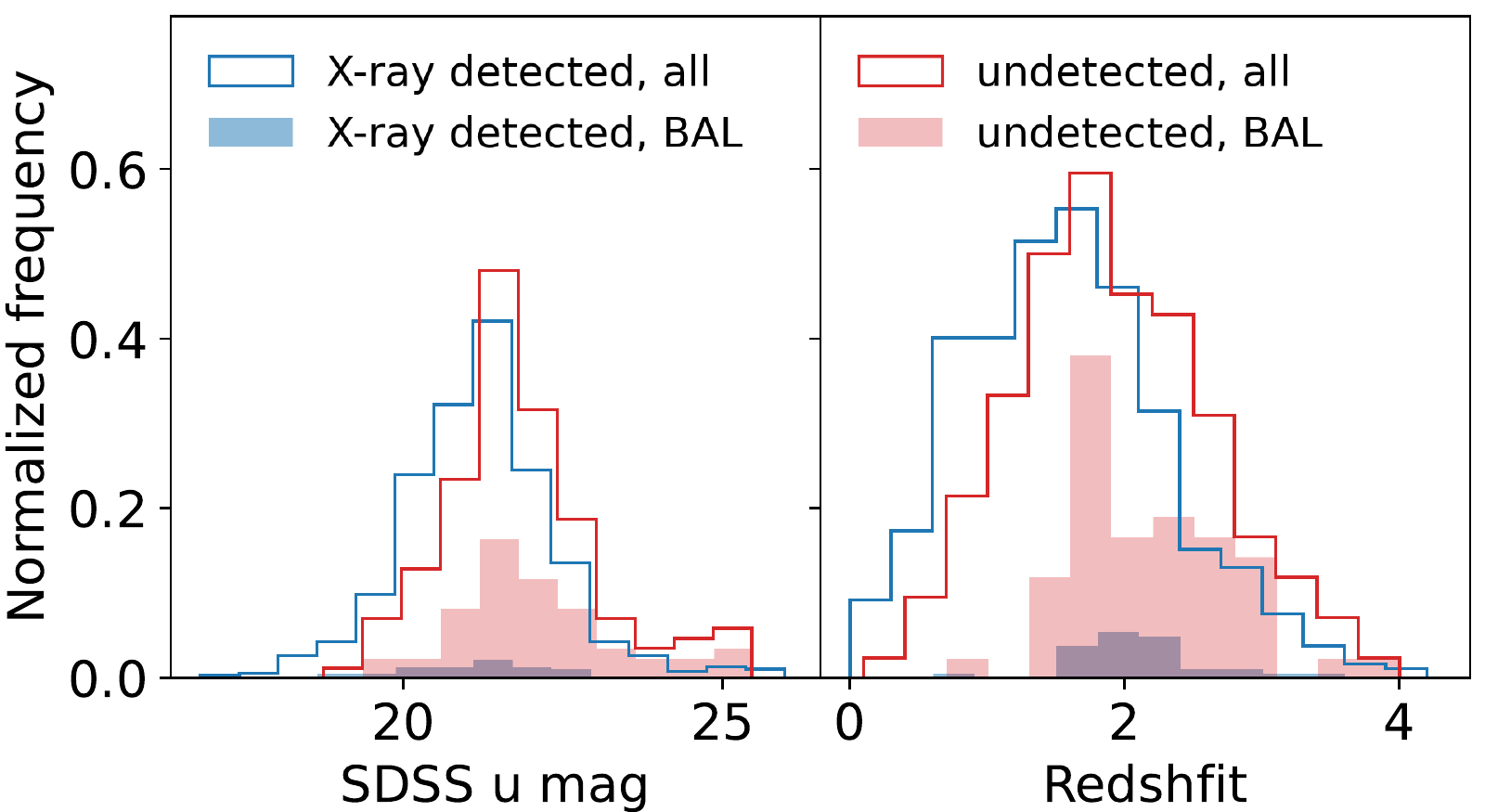}
\caption{Normalized $u$ magnitude (left) and redshift (right) distributions of the X-ray detected (blue) and undetected (red) SDSS-RM quasars.
  The shaded region shows the fraction of BAL quasars in each sample.
}
\label{fig:SDSSRM_det_und}
\end{figure}

  Among the $756$ SDSS-RM quasars (90\% of the whole sample) covered by the XMM exposures, $594$ (78\%) are identified as the best counterparts of X-ray sources.
  For the other $162$ quasars in the XMM FOV, it is not necessarily true that there are no X-ray signals at their positions. For $12$ of them, X-ray sources are detected within 16$\arcsec$ (the aperture radius used in this work).
$9$ out of the $12$ are identified as possible counterparts of X-ray sources but not the best ones.
We put the RMID of these $9$ quasars in the X-ray catalog, adding $1000$ to the RMID to separate them from the best counterparts.
Considering the uncertainties and complexities in both the X-ray positional accuracy and the multi-band counterpart association, we cannot rule out the possibility that they might be the true counterparts of the X-ray sources.

As described in \S~\ref{sec:poissonianlikelihood}, we perform forced photometry at the positions of all the $162$ quasars within a circle of radius 16$\arcsec$, and calculate aperture Poissonian detection likelihoods of them. We find $22$ quasars with a likelihood $>7$ in any of the three bands and consider them as X-ray detected, leaving $140$ quasars that are truly X-ray undetectable. 
The source net counts and upper limits are converted to fluxes using the EEF and Exp-ECF maps (Fig.~\ref{fig:EEF_ECF}) and to luminosities as described in \S~\ref{sec:SDSSspec}.
The luminosities and upper limits are plotted in Fig.~\ref{fig:L_z}.
These forced photometry results are also available with this paper as a supplementary catalog (see Appendix~\ref{app:Xraycolumns}).

Fig.~\ref{fig:SDSSRM_det_und} compares the X-ray detected SDSS-RM quasars, either as the best counterparts of the sources in the main XMM-RM catalog or through the Poissonian likelihood selection based on forced photometry, with the X-ray undetected ones.
These undetected quasars are relatively fainter (with a median $u$ magnitude of 21.7) and have relatively higher redshifts (median redshift 1.8) compared with the detected ones (median $u$ 21.2, median $z$ 1.5).
  Meanwhile, the fraction of BAL quasars, which are generally known to be X-ray weak \citep[e.g.][]{Gallagher2006,Gibson2009}, is much higher in the X-ray undetected sample (37\%) than in the X-ray detected sample (5\%).
  Excluding the $z<1.6$ part, where BAL identification is difficult because the broad CIV line is not well covered by the SDSS wavelength range, the fraction of BAL quasars in the X-ray undetected sample is 50\%.

\section{Discussion and Conclusion}
\subsection{The XMM-RM catalog}
With more than 40 \XMM{} observations that cover 90\% of the SDSS-RM field to an exposure depth of $\sim 15$ ks, the XMM-RM project adds to the comprehensive multi-wavelength coverage of the SDSS-RM quasar sample \citep{Shen2019}.
This paper presents the XMM-RM X-ray catalog and the optical/IR counterparts of the sources. More detailed analyses of X-ray spectra, multi-band colors, and spectral energy distributions based on this catalog will be presented in a subsequent paper.

We perform elaborate processing of the XMM data with the aim of optimizing the detection sensitivities for faint point sources, and perform source detection using a simultaneous PSF fitting technique which adopts the correct PSF model at any position of a camera during each individual observation.
We choose a relatively low PSF-fitting likelihood threshold of $L>3$, which corresponds to a spurious fraction of $\sim 5\%$, and compile a catalog of $3553$ sources.
We also calculate a Poissonian detection likelihood for each source, and perform further sample refinement on the basis of this likelihood.
Sub-samples selected in this manner show a logN-logS distribution consistent with previous X-ray surveys.

We combine the optical Legacy and the IR unWISE catalogs in the SDSS-RM field and search for counterparts of the X-ray sources in the combined catalog using the Bayesian method ``NWAY''.
We create a 2-dimensional, unWISE magnitude and color prior using Chandra catalogs, which have excellent positional accuracy. This prior is effective in improving the efficiency of counterpart identification.
Adopting $p_{any}>0.36$ and $p_{any}>0.63$ produces sub-samples of optical/IR counterparts of $2987$ (84\%) and $2648$ (74\%) sources, respectively.
According to Monte Carlo tests, the false association rates of these sub-samples are lower than $20\%$ and $10\%$, respectively.

We find SDSS DR16 spectra for $932$ of our X-ray sources. $831$ of them ($89\%$) are classified as AGN, and $594$ (71\%) of these AGN are in the SDSS-RM quasar catalog.
For the SDSS-RM quasars that are not associated with any X-ray sources, we calculate upper limits on their X-ray fluxes.

\subsection{X-ray catalogs: depth vs purity}

\begin{figure}[htbp]
\epsscale{1}
\plotone{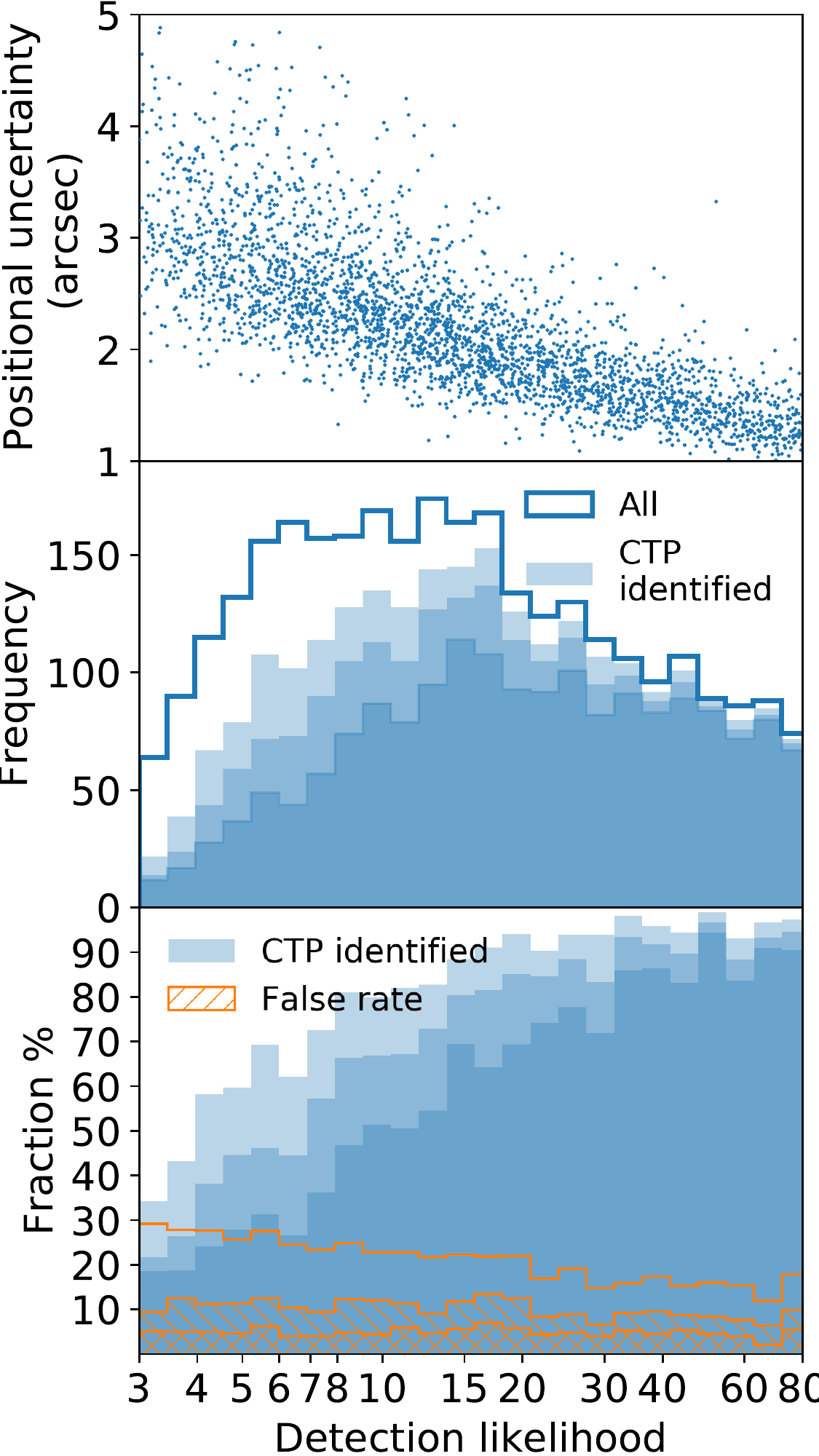}
\caption{The top panel is the scatter plot of the total positional uncertainty and PSF-fitting likelihood of the XMM-RM sources. The middle panel presents the likelihood distributions of all the XMM-RM sources (empty) and of the ones having counterparts (filled).
  The three levels of blue color depth of the filled histograms correspond to the $p_{any}$ thresholds of 0.36, 0.63, and 0.78, from light to dark; the corresponding fractions in each bin are displayed in the bottom panel.
  The orange histograms are the fractions of the randomly distributed sources with counterparts identified; the three levels of $p_{any}$ thresholds 0.36,0.63,and 0.78 are indicated with empty, loose shadows, and cross shadows, respectively.
}
\label{fig:ctpfraction}
\end{figure}

It has been common practice for X-ray surveys to provide high-purity catalogs with spurious fractions as low as $1\%$ \citep[e.g.,][]{Cappelluti2009,Laird2009,Georgakakis2011,Nandra2015,Civano2016,Chen2018}, prioritizing sample purity over survey depth.
In this work, we recommend a new approach dispensing with it, in consideration of the fact that these two figures of merit are preferred differently in different situations.
In the case of X-ray population studies, which rely on a well-defined X-ray selection function, it is essential to guarantee the sample purity through a high selection threshold, whereas a large number of sources at the faint end are expendable.
In the case of studying multi-band properties, especially when some sources detected in other bands are of special interest, it can be advantageous to lower the detection threshold to yield as many candidate X-ray sources as possible, even if many are spurious.
Therefore, we recommend to present a large master catalog adopting a relatively low detection likelihood.
Population analysis (e.g., logN-logS) of X-ray sources can be carried out by applying a post-hoc selection on the Poissonian detection likelihood, which gives rise to a high-purity sub-sample with a well-defined selection function. 
Such an approach will also be adopted in the on-going eROSITA X-ray surveys (\citealt{Merloni2012}, Brunner et al. in preparation).

Oriented towards the best completeness (or detection sensitivity),
we not only make every effort during the data reduction to increase the S/N, but also merge the catalogs detected in three bands, which yields a higher completeness at the expense of a poorly-defined X-ray selection function.
Choosing a low detection likelihood threshold of $L>3$, the XMM-RM catalog contains a large number of low-likelihood sources (one third at $L<10$; see the middle panel of Fig.~\ref{fig:ctpfraction}).
As discussed in \ref{sec:logNlogS}, the low detection likelihood is helpful as it improves the completeness of the Poissonian-likelihood selected sub-sample.
We also show the value of including these low-likelihood sources by identifying their optical/IR counterparts.
As shown in Fig.~\ref{fig:ctpfraction}, the counterpart identification completeness drops at low detection likelihoods.
It is not only caused by higher fractions of spurious sources at low likelihoods, but also because the low-likelihood sources
1) have larger positional uncertainties (Fig.~\ref{fig:ctpfraction} top panel) and thus lower posterior probabilities for real counterparts;
2) have their positions more easily affected by unaccounted factors, e.g., CCD gaps and nearby undetectable sources;
3) have relatively lower fluxes in both X-ray and optical/ IR bands and might drop out of the optical/IR detection limits.
The fraction of random-position counterparts (false rate) increases slightly with decreasing likelihood (Fig.~\ref{fig:ctpfraction} bottom panel), as a result of the increasing positional uncertainty.
Even when taking the increasing false rate into account, we have a significant fraction of sources at low likelihoods with reliable counterparts identified, at least above a likelihood of 4, not to mention that the false rate is overestimated (\S~\ref{sec:CTP}).
Considering the existence of multi-band counterparts as an additional prior, the posterior probabilities of these sources being real should be high. It would be a loss to exclude such sources, which are numerous, from the master catalog.

\acknowledgments
WNB acknowledges support from NASA grant 80NSSC19K0961 and NASA ADAP grant 80NSSC18K0878.
YS acknowledges support from an Alfred P. Sloan Research Fellowship and NSF grant AST-1715579.
LCH was supported by the National Science Foundation of China (11721303, 11991052) and the National Key R\&D Program of China (2016YFA0400702).

\begin{appendix}
\section{The XMM-RM catalogs}
\label{app:Xraycolumns}

  In addition to the primary XMM-RM catalog, we provide a supplementary catalog presenting the forced aperture photometry results of the X-ray undetected SDSS-RM quasars and the full NWAY output table matching the XMM-RM catalog to the combined Legacy-unWISE catalog adopting the unWISE magnitude-color prior, which contains all the possible counterparts of the X-ray sources. 
  They are available along with this paper and can also be obtained at \href{https://www.mpe.mpg.de/XraySurveys/XMM-RM/}{https://www.mpe.mpg.de/XraySurveys/XMM-RM/}.

  The columns of the primary XMM-RM catalog are described in three groups in \S~\ref{P1},\ref{P2}, and \ref{P3}.
In all the catalog, values are set to -99 if not applicable.

\subsection{Unique source parameters}
\label{P1}
  The following columns present basic information for each X-ray source.\\
ID: unique X-ray ID in the XMM-RM catalog\\
RA,DEC: X-ray coordinates (J2000) in degrees\\
RADEC\_ERR: X-ray positional uncertainty from PSF-fitting (combined RA--DEC error) in arcseconds\\
SYSERRCC: additional systematic positional uncertainty that needs to be added to RADEC\_ERR in quadrature to measure the total positional uncertainty\\
DetBand: the detection band is the band (F, S, H) that has the highest detection likelihood and in which the source position, DET\_ML, and extent are measured\\
DET\_ML: detection likelihood from PSF fitting using \texttt{emldetect}\\
EXT,EXT\_ERR: source extent and uncertainty (beta model core radius) in pixel (4$\arcsec$ per pixel)\\
EXT\_ML: source extent likelihood, i.e., the likelihood difference between a beta-model fit and a PSF-model fit\\
NHI: Galactic column density at the source position in cm$^{-2}$ from HI4PI \citep{HI4PI}\\
NNDist: distance to nearest neighbor in arcseconds, set to 120$\arcsec$ if having no neighbor within 120$\arcsec$\\
RMID: SDSS-RM catalog ID. RMID$>$1000 means the RMID is added by 1000 and this quasar is not identified as the best counterpart of the X-ray source. \\
SpecObjID: SDSS DR16 spectroscopic object ID \\
SDSS\_Cluster: ID in the \citet{Wen2012} SDSS cluster catalog\\
$z$,$zErr$: SDSS spectroscopic redshift, modified if necessary (\S~\ref{sec:SDSSspec}) \\
Class: SDSS spectroscopic classification (1: AGN; 2: galaxy; 3: star), modified if necessary (\S~\ref{sec:SDSSspec}) \\
Lx,LxErr: rest-frame 2-10 keV luminosity (erg s$^{-1}$) and 1-$\sigma$ uncertainty propagated form the flux uncertainty\\

\subsection{Parameters for each band}
\label{P2}

The following columns present the source properties measured in the full (F), soft (S), and hard (H) band. \\
SCTS\_[FSH],SCTS\_ERR\_[FSH]: source counts and error from PSF-fitting using \texttt{emldetect}\\
DET\_ML\_[FSH]: detection likelihood from PSF-fitting using \texttt{emldetect}\\
BG\_MAP\_[FSH]: background at source location in counts per pixel\\
RATE\_[FSH],RATE\_ERR\_[FSH]: source count rate and error in counts s$^{-1} $ from \texttt{emldetect}\\
RA\_[FSH],DEC\_[FSH]: coordinates (J2000) in degrees\\
RADEC\_ERR\_[FSH]: positional uncertainty from PSF-fitting (combined RA--DEC error) in arcseconds\\
EXP\_[FSH]: sum of the vignetted exposure (seconds) of the three EPIC cameras\\
Flux\_[FSH],FluxErr\_[FSH]: flux and error (10$^{-14}$ erg cm$^{-2}$ s$^{-1}$) in 0.5-10 (F), 0.5-2 (S), and 2-10 (H) keV band. In the primary catalog, they are measured from PSF-fitting count rate; in the supplementary forced-photometry catalog, they are measured from the aperture count rate. \\
NetCtsA\_[FSH]: background subtracted aperture source counts measured within 16" using eregionanalyse\\
TotCtsA\_[FSH]: total counts within 16" from eregionanalyse. TotCtsA - NetCtsA is the aperture background counts.\\
RateUpperA\_[FSH]: 2-$\sigma$ count rate upper limit on the background subtracted aperture source count rate .\\
EEF\_[FSH]: enclosed energy fraction of 16" at the source location on the EEF map\\
APLike\_[FSH]: aperture Poissonian likelihood, calculated adopting a maximum of 40.\\
ExpECF\_[FSH]: value at the source location on the Exp-ECF map, which is used to convert the net source counts in 0.5-7.5 (F), 0.5-2 (S), and 2-7.5 (H) keV band to fluxes in 0.5-10 (F), 0.5-2 (S), and 2-10 (H) band.\\

\subsection{Parameters of counterparts}
\label{P3}

  The following columns present NWAY association information of the best optical-IR counterpart from the combined Legacy--unWISE catalog. Please refer to \citet{Salvato2018} for more detailed descriptions.\\
UW\_RA,UW\_DEC: unWISE source coordinates (J2000) in degrees\\
UW\_ID: unWISE source ID\\
UW\_W1\_AB,UW\_W1ERR: unWISE W1 AB-magnitude and error\\
UW\_W2\_AB,UW\_W2ERR: unWISE W2 AB-magnitude and error\\
LS\_RA,LS\_DEC: Legacy source coordinates (J2000) in degrees\\
LS\_BRICKNAME: Legacy brick name\\
LS\_BRICKID: Legacy brick ID\\
LS\_OBJID: Legacy object ID\\
LS\_Mag\_[rgz]: Legacy AB-magnitude in the r,g,and z band.\\
$p_{any}$: the probability that any of the associations is the correct one\\
$p_i$: relative probability of the match\\
match\_flag: 1 for the most probable match, 2 for less probable matches with $p_i/p_i^{best} > 0.5$\\
Separation: separation between the pair of sources in arcsec\\
dist\_bayesfactor: logarithm of ratio between prior and posterior from distance matching\\
dist\_post: distance posterior probability comparing this association versus no association\\
$p_{single}$: same as dist\_post, but weighted by the unWISE magnitude-color prior\\
CellInd: index of the $W1+W2 \sim 4(W1-W2)$ cell in which the source is located (\S~\ref{sec:colorprior})\\
bias\_UWLS\_CellInd: probability weighting introduced by the unWISE color prior\\

\subsection{Exclusive columns of the SDSS-RM forced-photometry catalog}

  Most of the columns of the supplementary catalog are the same as the primary catalog (\S~\ref{P1},\ref{P2}), except a few ones:\\
RA,DEC: the SDSS-RM quasar coordinates (J2000) in degrees\\
FluxUpper\_[FSH]: 2-$\sigma$ flux upper limit measured from RateUpperA\\
LxUpper: 2-$\sigma$ upper limit of rest-frame 2-10 keV luminosity (erg s$^{-1}$) measured from RateUpperA\\

\subsection{Exclusive columns of the NWAY output table}

  Most of the columns of the NWAY output table are already described in \S~\ref{P3}, except the following ones:\\
X\_ID,X\_RA,X\_DEC,X\_POSERR: the X-ray ID, coordinates, and total positional uncertainties (\S~\ref{P1})\\
UWLS\_RA,UWLS\_DEC,UWLS\_POSERR: the coordinates and positional uncertainties of the combined Legacy-unWISE catalog, adopting that of the Legacy sources when available and the unWISE sources in the other cases\\

\end{appendix}

\bibliography{rm}

\begin{thebibliography}{}
\expandafter\ifx\csname natexlab\endcsname\relax\def\natexlab#1{#1}\fi
\providecommand{\url}[1]{\href{#1}{#1}}
\providecommand{\dodoi}[1]{doi:~\href{http://doi.org/#1}{\nolinkurl{#1}}}
\providecommand{\doeprint}[1]{\href{http://ascl.net/#1}{\nolinkurl{http://ascl.net/#1}}}
\providecommand{\doarXiv}[1]{\href{https://arxiv.org/abs/#1}{\nolinkurl{https://arxiv.org/abs/#1}}}

\bibitem[{{Blanton} {et~al.}(2017){Blanton}, {Bershady}, {Abolfathi},
  {Albareti}, {Allende Prieto}, {Almeida}, {Alonso-Garc{\'\i}a}, {Anders},
  {Anderson}, {Andrews}, {Aquino-Ort{\'\i}z}, {Arag{\'o}n-Salamanca},
  {Argudo-Fern{\'a}ndez}, {Armengaud}, {Aubourg}, {Avila-Reese}, {Badenes},
  {Bailey}, {Barger}, {Barrera-Ballesteros}, {Bartosz}, {Bates}, {Baumgarten},
  {Bautista}, {Beaton}, {Beers}, {Belfiore}, {Bender}, {Berlind}, {Bernardi},
  {Beutler}, {Bird}, {Bizyaev}, {Blanc}, {Blomqvist}, {Bolton}, {Boquien},
  {Borissova}, {van den Bosch}, {Bovy}, {Brandt}, {Brinkmann}, {Brownstein},
  {Bundy}, {Burgasser}, {Burtin}, {Busca}, {Cappellari}, {Delgado Carigi},
  {Carlberg}, {Carnero Rosell}, {Carrera}, {Chanover}, {Cherinka}, {Cheung},
  {G{\'o}mez Maqueo Chew}, {Chiappini}, {Choi}, {Chojnowski}, {Chuang},
  {Chung}, {Cirolini}, {Clerc}, {Cohen}, {Comparat}, {da Costa}, {Cousinou},
  {Covey}, {Crane}, {Croft}, {Cruz-Gonzalez}, {Garrido Cuadra}, {Cunha},
  {Damke}, {Darling}, {Davies}, {Dawson}, {de la Macorra}, {Dell'Agli}, {De
  Lee}, {Delubac}, {Di Mille}, {Diamond-Stanic}, {Cano-D{\'\i}az}, {Donor},
  {Downes}, {Drory}, {du Mas des Bourboux}, {Duckworth}, {Dwelly}, {Dyer},
  {Ebelke}, {Eigenbrot}, {Eisenstein}, {Emsellem}, {Eracleous}, {Escoffier},
  {Evans}, {Fan}, {Fern{\'a}ndez-Alvar}, {Fernandez-Trincado}, {Feuillet},
  {Finoguenov}, {Fleming}, {Font-Ribera}, {Fredrickson}, {Freischlad},
  {Frinchaboy}, {Fuentes}, {Galbany}, {Garcia-Dias},
  {Garc{\'\i}a-Hern{\'a}ndez}, {Gaulme}, {Geisler}, {Gelfand},
  {Gil-Mar{\'\i}n}, {Gillespie}, {Goddard}, {Gonzalez-Perez}, {Grabowski},
  {Green}, {Grier}, {Gunn}, {Guo}, {Guy}, {Hagen}, {Hahn}, {Hall}, {Harding},
  {Hasselquist}, {Hawley}, {Hearty}, {Gonzalez Hern{\'a}ndez}, {Ho}, {Hogg},
  {Holley-Bockelmann}, {Holtzman}, {Holzer}, {Huehnerhoff}, {Hutchinson},
  {Hwang}, {Ibarra-Medel}, {da Silva Ilha}, {Ivans}, {Ivory}, {Jackson},
  {Jensen}, {Johnson}, {Jones}, {J{\"o}nsson}, {Jullo}, {Kamble}, {Kinemuchi},
  {Kirkby}, {Kitaura}, {Klaene}, {Knapp}, {Kneib}, {Kollmeier}, {Lacerna},
  {Lane}, {Lang}, {Law}, {Lazarz}, {Lee}, {Le Goff}, {Liang}, {Li}, {Li},
  {Lian}, {Lima}, {Lin}, {Lin}, {Bertran de Lis}, {Liu}, {de Icaza Lizaola},
  {Long}, {Lucatello}, {Lundgren}, {MacDonald}, {Deconto Machado}, {MacLeod},
  {Mahadevan}, {Geimba Maia}, {Maiolino}, {Majewski}, {Malanushenko},
  {Malanushenko}, {Manchado}, {Mao}, {Maraston}, {Marques-Chaves}, {Masseron},
  {Masters}, {McBride}, {McDermid}, {McGrath}, {McGreer}, {Medina Pe{\~n}a},
  {Melendez}, {Merloni}, {Merrifield}, {Meszaros}, {Meza}, {Minchev},
  {Minniti}, {Miyaji}, {More}, {Mulchaey}, {M{\"u}ller-S{\'a}nchez}, {Muna},
  {Munoz}, {Myers}, {Nair}, {Nandra}, {Correa do Nascimento}, {Negrete},
  {Ness}, {Newman}, {Nichol}, {Nidever}, {Nitschelm}, {Ntelis}, {O'Connell},
  {Oelkers}, {Oravetz}, {Oravetz}, {Pace}, {Padilla}, {Palanque-Delabrouille},
  {Alonso Palicio}, {Pan}, {Parejko}, {Parikh}, {P{\^a}ris}, {Park}, {Patten},
  {Peirani}, {Pellejero-Ibanez}, {Penny}, {Percival}, {Perez-Fournon},
  {Petitjean}, {Pieri}, {Pinsonneault}, {Pisani}, {Poleski}, {Prada},
  {Prakash}, {Queiroz}, {Raddick}, {Raichoor}, {Barboza Rembold}, {Richstein},
  {Riffel}, {Riffel}, {Rix}, {Robin}, {Rockosi}, {Rodr{\'\i}guez-Torres},
  {Roman-Lopes}, {Rom{\'a}n-Z{\'u}{\~n}iga}, {Rosado}, {Ross}, {Rossi}, {Ruan},
  {Ruggeri}, {Rykoff}, {Salazar-Albornoz}, {Salvato}, {S{\'a}nchez}, {Aguado},
  {S{\'a}nchez-Gallego}, {Santana}, {Santiago}, {Sayres}, {Schiavon}, {da Silva
  Schimoia}, {Schlafly}, {Schlegel}, {Schneider}, {Schultheis}, {Schuster},
  {Schwope}, {Seo}, {Shao}, {Shen}, {Shetrone}, {Shull}, {Simon}, {Skinner},
  {Skrutskie}, {Slosar}, {Smith}, {Sobeck}, {Sobreira}, {Somers}, {Souto},
  {Stark}, {Stassun}, {Stauffer}, {Steinmetz}, {Storchi-Bergmann},
  {Streblyanska}, {Stringfellow}, {Su{\'a}rez}, {Sun}, {Suzuki}, {Szigeti},
  {Taghizadeh-Popp}, {Tang}, {Tao}, {Tayar}, {Tembe}, {Teske}, {Thakar},
  {Thomas}, {Thompson}, {Tinker}, {Tissera}, {Tojeiro}, {Hernandez Toledo}, {de
  la Torre}, {Tremonti}, {Troup}, {Valenzuela}, {Martinez Valpuesta},
  {Vargas-Gonz{\'a}lez}, {Vargas-Maga{\~n}a}, {Vazquez}, {Villanova}, {Vivek},
  {Vogt}, {Wake}, {Walterbos}, {Wang}, {Weaver}, {Weijmans}, {Weinberg},
  {Westfall}, {Whelan}, {Wild}, {Wilson}, {Wood-Vasey}, {Wylezalek}, {Xiao},
  {Yan}, {Yang}, {Ybarra}, {Y{\`e}che}, {Zakamska}, {Zamora}, {Zarrouk},
  {Zasowski}, {Zhang}, {Zhao}, {Zheng}, {Zheng}, {Zhou}, {Zhou}, {Zhu},
  {Zoccali}, \& {Zou}}]{Blaton2017}
{Blanton}, M.~R., {Bershady}, M.~A., {Abolfathi}, B., {et~al.} 2017, \aj, 154,
  28, \dodoi{10.3847/1538-3881/aa7567}

\bibitem[{{Cappelluti} {et~al.}(2009){Cappelluti}, {Brusa}, {Hasinger},
  {Comastri}, {Zamorani}, {Finoguenov}, {Gilli}, {Puccetti}, {Miyaji},
  {Salvato}, {Vignali}, {Aldcroft}, {B{\"o}hringer}, {Brunner}, {Civano},
  {Elvis}, {Fiore}, {Fruscione}, {Griffiths}, {Guzzo}, {Iovino}, {Koekemoer},
  {Mainieri}, {Scoville}, {Shopbell}, {Silverman}, \& {Urry}}]{Cappelluti2009}
{Cappelluti}, N., {Brusa}, M., {Hasinger}, G., {et~al.} 2009, \aap, 497, 635,
  \dodoi{10.1051/0004-6361/200810794}

\bibitem[{{Cash}(1979)}]{Cash1979}
{Cash}, W. 1979, \apj, 228, 939, \dodoi{10.1086/156922}

\bibitem[{{Chen} {et~al.}(2018){Chen}, {Brandt}, {Luo}, {Ranalli}, {Yang},
  {Alexander}, {Bauer}, {Kelson}, {Lacy}, {Nyland}, {Tozzi}, {Vito},
  {Cirasuolo}, {Gilli}, {Jarvis}, {Lehmer}, {Paolillo}, {Schneider}, {Shemmer},
  {Smail}, {Sun}, {Tanaka}, {Vaccari}, {Vignali}, {Xue}, {Banerji}, {Chow},
  {H{\"a}u{\ss}ler}, {Norris}, {Silverman}, \& {Trump}}]{Chen2018}
{Chen}, C. T.~J., {Brandt}, W.~N., {Luo}, B., {et~al.} 2018, \mnras, 478, 2132,
  \dodoi{10.1093/mnras/sty1036}

\bibitem[{{Civano} {et~al.}(2016){Civano}, {Marchesi}, {Comastri}, {Urry},
  {Elvis}, {Cappelluti}, {Puccetti}, {Brusa}, {Zamorani}, {Hasinger},
  {Aldcroft}, {Alexand er}, {Allevato}, {Brunner}, {Capak}, {Finoguenov},
  {Fiore}, {Fruscione}, {Gilli}, {Glotfelty}, {Griffiths}, {Hao}, {Harrison},
  {Jahnke}, {Kartaltepe}, {Karim}, {LaMassa}, {Lanzuisi}, {Miyaji}, {Ranalli},
  {Salvato}, {Sargent}, {Scoville}, {Schawinski}, {Schinnerer}, {Silverman},
  {Smolcic}, {Stern}, {Toft}, {Trakhtenbrot}, {Treister}, \&
  {Vignali}}]{Civano2016}
{Civano}, F., {Marchesi}, S., {Comastri}, A., {et~al.} 2016, \apj, 819, 62,
  \dodoi{10.3847/0004-637X/819/1/62}

\bibitem[{{Davis} {et~al.}(2007){Davis}, {Guhathakurta}, {Konidaris}, {Newman},
  {Ashby}, {Biggs}, {Barmby}, {Bundy}, {Chapman}, {Coil}, {Conselice},
  {Cooper}, {Croton}, {Eisenhardt}, {Ellis}, {Faber}, {Fang}, {Fazio},
  {Georgakakis}, {Gerke}, {Goss}, {Gwyn}, {Harker}, {Hopkins}, {Huang},
  {Ivison}, {Kassin}, {Kirby}, {Koekemoer}, {Koo}, {Laird}, {Le Floc'h}, {Lin},
  {Lotz}, {Marshall}, {Martin}, {Metevier}, {Moustakas}, {Nandra}, {Noeske},
  {Papovich}, {Phillips}, {Rich}, {Rieke}, {Rigopoulou}, {Salim},
  {Schiminovich}, {Simard}, {Smail}, {Small}, {Weiner}, {Willmer}, {Willner},
  {Wilson}, {Wright}, \& {Yan}}]{Davis2007}
{Davis}, M., {Guhathakurta}, P., {Konidaris}, N.~P., {et~al.} 2007, \apjl, 660,
  L1, \dodoi{10.1086/517931}

\bibitem[{{Dawson} {et~al.}(2013){Dawson}, {Schlegel}, {Ahn}, {Anderson},
  {Aubourg}, {Bailey}, {Barkhouser}, {Bautista}, {Beifiori}, {Berlind},
  {Bhardwaj}, {Bizyaev}, {Blake}, {Blanton}, {Blomqvist}, {Bolton}, {Borde},
  {Bovy}, {Brandt}, {Brewington}, {Brinkmann}, {Brown}, {Brownstein}, {Bundy},
  {Busca}, {Carithers}, {Carnero}, {Carr}, {Chen}, {Comparat}, {Connolly},
  {Cope}, {Croft}, {Cuesta}, {da Costa}, {Davenport}, {Delubac}, {de Putter},
  {Dhital}, {Ealet}, {Ebelke}, {Eisenstein}, {Escoffier}, {Fan}, {Filiz Ak},
  {Finley}, {Font-Ribera}, {G{\'e}nova-Santos}, {Gunn}, {Guo}, {Haggard},
  {Hall}, {Hamilton}, {Harris}, {Harris}, {Ho}, {Hogg}, {Holder}, {Honscheid},
  {Huehnerhoff}, {Jordan}, {Jordan}, {Kauffmann}, {Kazin}, {Kirkby}, {Klaene},
  {Kneib}, {Le Goff}, {Lee}, {Long}, {Loomis}, {Lundgren}, {Lupton}, {Maia},
  {Makler}, {Malanushenko}, {Malanushenko}, {Mandelbaum}, {Manera}, {Maraston},
  {Margala}, {Masters}, {McBride}, {McDonald}, {McGreer}, {McMahon}, {Mena},
  {Miralda-Escud{\'e}}, {Montero-Dorta}, {Montesano}, {Muna}, {Myers},
  {Naugle}, {Nichol}, {Noterdaeme}, {Nuza}, {Olmstead}, {Oravetz}, {Oravetz},
  {Owen}, {Padmanabhan}, {Palanque-Delabrouille}, {Pan}, {Parejko},
  {P{\^a}ris}, {Percival}, {P{\'e}rez-Fournon}, {P{\'e}rez-R{\`a}fols},
  {Petitjean}, {Pfaffenberger}, {Pforr}, {Pieri}, {Prada}, {Price-Whelan},
  {Raddick}, {Rebolo}, {Rich}, {Richards}, {Rockosi}, {Roe}, {Ross}, {Ross},
  {Rossi}, {Rubi{\~n}o-Martin}, {Samushia}, {S{\'a}nchez}, {Sayres}, {Schmidt},
  {Schneider}, {Sc{\'o}ccola}, {Seo}, {Shelden}, {Sheldon}, {Shen}, {Shu},
  {Slosar}, {Smee}, {Snedden}, {Stauffer}, {Steele}, {Strauss}, {Streblyanska},
  {Suzuki}, {Swanson}, {Tal}, {Tanaka}, {Thomas}, {Tinker}, {Tojeiro},
  {Tremonti}, {Vargas Maga{\~n}a}, {Verde}, {Viel}, {Wake}, {Watson}, {Weaver},
  {Weinberg}, {Weiner}, {West}, {White}, {Wood-Vasey}, {Yeche}, {Zehavi},
  {Zhao}, \& {Zheng}}]{Dawson2013}
{Dawson}, K.~S., {Schlegel}, D.~J., {Ahn}, C.~P., {et~al.} 2013, \aj, 145, 10,
  \dodoi{10.1088/0004-6256/145/1/10}

\bibitem[{{Dey} {et~al.}(2019){Dey}, {Schlegel}, {Lang}, {Blum}, {Burleigh},
  {Fan}, {Findlay}, {Finkbeiner}, {Herrera}, {Juneau}, {Landriau}, {Levi},
  {McGreer}, {Meisner}, {Myers}, {Moustakas}, {Nugent}, {Patej}, {Schlafly},
  {Walker}, {Valdes}, {Weaver}, {Y{\`e}che}, {Zou}, {Zhou}, {Abareshi},
  {Abbott}, {Abolfathi}, {Aguilera}, {Alam}, {Allen}, {Alvarez}, {Annis},
  {Ansarinejad}, {Aubert}, {Beechert}, {Bell}, {BenZvi}, {Beutler}, {Bielby},
  {Bolton}, {Brice{\~n}o}, {Buckley-Geer}, {Butler}, {Calamida}, {Carlberg},
  {Carter}, {Casas}, {Castander}, {Choi}, {Comparat}, {Cukanovaite}, {Delubac},
  {DeVries}, {Dey}, {Dhungana}, {Dickinson}, {Ding}, {Donaldson}, {Duan},
  {Duckworth}, {Eftekharzadeh}, {Eisenstein}, {Etourneau}, {Fagrelius},
  {Farihi}, {Fitzpatrick}, {Font-Ribera}, {Fulmer}, {G{\"a}nsicke},
  {Gaztanaga}, {George}, {Gerdes}, {Gontcho}, {Gorgoni}, {Green}, {Guy},
  {Harmer}, {Hernand ez}, {Honscheid}, {Huang}, {James}, {Jannuzi}, {Jiang},
  {Joyce}, {Karcher}, {Karkar}, {Kehoe}, {Kneib}, {Kueter-Young}, {Lan},
  {Lauer}, {Le Guillou}, {Le Van Suu}, {Lee}, {Lesser}, {Perreault Levasseur},
  {Li}, {Mann}, {Marshall}, {Mart{\'\i}nez-V{\'a}zquez}, {Martini}, {du Mas des
  Bourboux}, {McManus}, {Meier}, {M{\'e}nard}, {Metcalfe},
  {Mu{\~n}oz-Guti{\'e}rrez}, {Najita}, {Napier}, {Narayan}, {Newman}, {Nie},
  {Nord}, {Norman}, {Olsen}, {Paat}, {Palanque-Delabrouille}, {Peng},
  {Poppett}, {Poremba}, {Prakash}, {Rabinowitz}, {Raichoor}, {Rezaie},
  {Robertson}, {Roe}, {Ross}, {Ross}, {Rudnick}, {Safonova}, {Saha},
  {S{\'a}nchez}, {Savary}, {Schweiker}, {Scott}, {Seo}, {Shan}, {Silva},
  {Slepian}, {Soto}, {Sprayberry}, {Staten}, {Stillman}, {Stupak}, {Summers},
  {Sien Tie}, {Tirado}, {Vargas-Maga{\~n}a}, {Vivas}, {Wechsler}, {Williams},
  {Yang}, {Yang}, {Yapici}, {Zaritsky}, {Zenteno}, {Zhang}, {Zhang}, {Zhou}, \&
  {Zhou}}]{Dey2019}
{Dey}, A., {Schlegel}, D.~J., {Lang}, D., {et~al.} 2019, \aj, 157, 168,
  \dodoi{10.3847/1538-3881/ab089d}

\bibitem[{{Eisenstein} {et~al.}(2011){Eisenstein}, {Weinberg}, {Agol},
  {Aihara}, {Allende Prieto}, {Anderson}, {Arns}, {Aubourg}, {Bailey},
  {Balbinot}, {Barkhouser}, {Beers}, {Berlind}, {Bickerton}, {Bizyaev},
  {Blanton}, {Bochanski}, {Bolton}, {Bosman}, {Bovy}, {Brandt}, {Breslauer},
  {Brewington}, {Brinkmann}, {Brown}, {Brownstein}, {Burger}, {Busca},
  {Campbell}, {Cargile}, {Carithers}, {Carlberg}, {Carr}, {Chang}, {Chen},
  {Chiappini}, {Comparat}, {Connolly}, {Cortes}, {Croft}, {Cunha}, {da Costa},
  {Davenport}, {Dawson}, {De Lee}, {Porto de Mello}, {de Simoni}, {Dean},
  {Dhital}, {Ealet}, {Ebelke}, {Edmondson}, {Eiting}, {Escoffier}, {Esposito},
  {Evans}, {Fan}, {Femen{\'\i}a Castell{\'a}}, {Dutra Ferreira}, {Fitzgerald},
  {Fleming}, {Font-Ribera}, {Ford}, {Frinchaboy}, {Garc{\'\i}a P{\'e}rez},
  {Gaudi}, {Ge}, {Ghezzi}, {Gillespie}, {Gilmore}, {Girardi}, {Gott}, {Gould},
  {Grebel}, {Gunn}, {Hamilton}, {Harding}, {Harris}, {Hawley}, {Hearty},
  {Hennawi}, {Gonz{\'a}lez Hern{\'a}ndez}, {Ho}, {Hogg}, {Holtzman},
  {Honscheid}, {Inada}, {Ivans}, {Jiang}, {Jiang}, {Johnson}, {Jordan},
  {Jordan}, {Kauffmann}, {Kazin}, {Kirkby}, {Klaene}, {Knapp}, {Kneib},
  {Kochanek}, {Koesterke}, {Kollmeier}, {Kron}, {Lampeitl}, {Lang}, {Lawler},
  {Le Goff}, {Lee}, {Lee}, {Leisenring}, {Lin}, {Liu}, {Long}, {Loomis},
  {Lucatello}, {Lundgren}, {Lupton}, {Ma}, {Ma}, {MacDonald}, {Mack},
  {Mahadevan}, {Maia}, {Majewski}, {Makler}, {Malanushenko}, {Malanushenko},
  {Mand elbaum}, {Maraston}, {Margala}, {Maseman}, {Masters}, {McBride},
  {McDonald}, {McGreer}, {McMahon}, {Mena Requejo}, {M{\'e}nard},
  {Miralda-Escud{\'e}}, {Morrison}, {Mullally}, {Muna}, {Murayama}, {Myers},
  {Naugle}, {Neto}, {Nguyen}, {Nichol}, {Nidever}, {O'Connell}, {Ogando},
  {Olmstead}, {Oravetz}, {Padmanabhan}, {Paegert}, {Palanque-Delabrouille},
  {Pan}, {Pandey}, {Parejko}, {P{\^a}ris}, {Pellegrini}, {Pepper}, {Percival},
  {Petitjean}, {Pfaffenberger}, {Pforr}, {Phleps}, {Pichon}, {Pieri}, {Prada},
  {Price-Whelan}, {Raddick}, {Ramos}, {Reid}, {Reyle}, {Rich}, {Richards},
  {Rieke}, {Rieke}, {Rix}, {Robin}, {Rocha-Pinto}, {Rockosi}, {Roe},
  {Rollinde}, {Ross}, {Ross}, {Rossetto}, {S{\'a}nchez}, {Santiago}, {Sayres},
  {Schiavon}, {Schlegel}, {Schlesinger}, {Schmidt}, {Schneider}, {Sellgren},
  {Shelden}, {Sheldon}, {Shetrone}, {Shu}, {Silverman}, {Simmerer}, {Simmons},
  {Sivarani}, {Skrutskie}, {Slosar}, {Smee}, {Smith}, {Snedden}, {Stassun},
  {Steele}, {Steinmetz}, {Stockett}, {Stollberg}, {Strauss}, {Szalay},
  {Tanaka}, {Thakar}, {Thomas}, {Tinker}, {Tofflemire}, {Tojeiro}, {Tremonti},
  {Vargas Maga{\~n}a}, {Verde}, {Vogt}, {Wake}, {Wan}, {Wang}, {Weaver},
  {White}, {White}, {Wilson}, {Wisniewski}, {Wood-Vasey}, {Yanny}, {Yasuda},
  {Y{\`e}che}, {York}, {Young}, {Zasowski}, {Zehavi}, \&
  {Zhao}}]{Eisenstein2011}
{Eisenstein}, D.~J., {Weinberg}, D.~H., {Agol}, E., {et~al.} 2011, \aj, 142,
  72, \dodoi{10.1088/0004-6256/142/3/72}

\bibitem[{{Gabriel} {et~al.}(2004){Gabriel}, {Denby}, {Fyfe}, {Hoar}, {Ibarra},
  {Ojero}, {Osborne}, {Saxton}, {Lammers}, \& {Vacanti}}]{Gabriel2004}
{Gabriel}, C., {Denby}, M., {Fyfe}, D.~J., {et~al.} 2004, in Astronomical
  Society of the Pacific Conference Series, Vol. 314, Astronomical Data
  Analysis Software and Systems (ADASS) XIII, ed. F.~{Ochsenbein}, M.~G.
  {Allen}, \& D.~{Egret}, 759

\bibitem[{{Gallagher} {et~al.}(2006){Gallagher}, {Brandt}, {Chartas},
  {Priddey}, {Garmire}, \& {Sambruna}}]{Gallagher2006}
{Gallagher}, S.~C., {Brandt}, W.~N., {Chartas}, G., {et~al.} 2006, \apj, 644,
  709, \dodoi{10.1086/503762}

\bibitem[{{Georgakakis} \& {Nandra}(2011)}]{Georgakakis2011}
{Georgakakis}, A., \& {Nandra}, K. 2011, \mnras, 414, 992,
  \dodoi{10.1111/j.1365-2966.2011.18387.x}

\bibitem[{{Georgakakis} {et~al.}(2008){Georgakakis}, {Nandra}, {Laird}, {Aird},
  \& {Trichas}}]{Georgakakis2008}
{Georgakakis}, A., {Nandra}, K., {Laird}, E.~S., {Aird}, J., \& {Trichas}, M.
  2008, \mnras, 388, 1205, \dodoi{10.1111/j.1365-2966.2008.13423.x}

\bibitem[{{Gezari} {et~al.}(2013){Gezari}, {Martin}, {Forster}, {Neill},
  {Huber}, {Heckman}, {Bianchi}, {Morrissey}, {Neff}, {Seibert},
  {Schiminovich}, {Wyder}, {Burgett}, {Chambers}, {Kaiser}, {Magnier}, {Price},
  \& {Tonry}}]{Gezari2013}
{Gezari}, S., {Martin}, D.~C., {Forster}, K., {et~al.} 2013, \apj, 766, 60,
  \dodoi{10.1088/0004-637X/766/1/60}

\bibitem[{{Gibson} {et~al.}(2009){Gibson}, {Jiang}, {Brandt}, {Hall}, {Shen},
  {Wu}, {Anderson}, {Schneider}, {Vand en Berk}, {Gallagher}, {Fan}, \&
  {York}}]{Gibson2009}
{Gibson}, R.~R., {Jiang}, L., {Brandt}, W.~N., {et~al.} 2009, \apj, 692, 758,
  \dodoi{10.1088/0004-637X/692/1/758}

\bibitem[{{Goulding} {et~al.}(2012){Goulding}, {Forman}, {Hickox}, {Jones},
  {Kraft}, {Murray}, {Vikhlinin}, {Coil}, {Cooper}, {Davis}, \&
  {Newman}}]{Goulding2012}
{Goulding}, A.~D., {Forman}, W.~R., {Hickox}, R.~C., {et~al.} 2012, \apjs, 202,
  6, \dodoi{10.1088/0067-0049/202/1/6}

\bibitem[{{HI4PI Collaboration} {et~al.}(2016){HI4PI Collaboration}, {Ben
  Bekhti}, {Fl{\"o}er}, {Keller}, {Kerp}, {Lenz}, {Winkel}, {Bailin},
  {Calabretta}, {Dedes}, {Ford}, {Gibson}, {Haud}, {Janowiecki}, {Kalberla},
  {Lockman}, {McClure-Griffiths}, {Murphy}, {Nakanishi}, {Pisano}, \&
  {Staveley-Smith}}]{HI4PI}
{HI4PI Collaboration}, {Ben Bekhti}, N., {Fl{\"o}er}, L., {et~al.} 2016, \aap,
  594, A116, \dodoi{10.1051/0004-6361/201629178}

\bibitem[{{Kaiser} {et~al.}(2010){Kaiser}, {Burgett}, {Chambers}, {Denneau},
  {Heasley}, {Jedicke}, {Magnier}, {Morgan}, {Onaka}, \& {Tonry}}]{Kaiser2010}
{Kaiser}, N., {Burgett}, W., {Chambers}, K., {et~al.} 2010, 7733, 77330E,
  \dodoi{10.1117/12.859188}

\bibitem[{Komatsu {et~al.}(2011)Komatsu, Smith, Dunkley, Bennett, Gold,
  Hinshaw, Jarosik, Larson, Nolta, Page, Spergel, Halpern, Hill, Kogut, Limon,
  Meyer, Odegard, Tucker, Weiland, Wollack, \& Wright}]{Komatsu2011}
Komatsu, E., Smith, K.~M., Dunkley, J., {et~al.} 2011, The Astrophysical
  Journal Supplement Series, 192, 18, \dodoi{10.1088/0067-0049/192/2/18}

\bibitem[{{Laird} {et~al.}(2009){Laird}, {Nandra}, {Georgakakis}, {Aird},
  {Barmby}, {Conselice}, {Coil}, {Davis}, {Faber}, {Fazio}, {Guhathakurta},
  {Koo}, {Sarajedini}, \& {Willmer}}]{Laird2009}
{Laird}, E.~S., {Nandra}, K., {Georgakakis}, A., {et~al.} 2009, VizieR Online
  Data Catalog, J/ApJS/180/102

\bibitem[{{LaMassa} {et~al.}(2016){LaMassa}, {Urry}, {Cappelluti},
  {B{\"o}hringer}, {Comastri}, {Glikman}, {Richards}, {Ananna}, {Brusa},
  {Cardamone}, {Chon}, {Civano}, {Farrah}, {Gilfanov}, {Green}, {Komossa},
  {Lira}, {Makler}, {Marchesi}, {Pecoraro}, {Ranalli}, {Salvato}, {Schawinski},
  {Stern}, {Treister}, \& {Viero}}]{LaMassa2016}
{LaMassa}, S.~M., {Urry}, C.~M., {Cappelluti}, N., {et~al.} 2016, \apj, 817,
  172, \dodoi{10.3847/0004-637X/817/2/172}

\bibitem[{{Lang} {et~al.}(2016){Lang}, {Hogg}, \& {Schlegel}}]{Lang2016}
{Lang}, D., {Hogg}, D.~W., \& {Schlegel}, D.~J. 2016, \aj, 151, 36,
  \dodoi{10.3847/0004-6256/151/2/36}

\bibitem[{{Liu} {et~al.}(2013){Liu}, {Tozzi}, {Tundo}, {Moretti}, {Wang},
  {Rosati}, \& {Guglielmetti}}]{Liu2013}
{Liu}, T., {Tozzi}, P., {Tundo}, E., {et~al.} 2013, \aap, 549, A143,
  \dodoi{10.1051/0004-6361/201219866}

\bibitem[{{Luo} {et~al.}(2017){Luo}, {Brandt}, {Xue}, {Lehmer}, {Alexander},
  {Bauer}, {Vito}, {Yang}, {Basu-Zych}, {Comastri}, {Gilli}, {Gu},
  {Hornschemeier}, {Koekemoer}, {Liu}, {Mainieri}, {Paolillo}, {Ranalli},
  {Rosati}, {Schneider}, {Shemmer}, {Smail}, {Sun}, {Tozzi}, {Vignali}, \&
  {Wang}}]{Luo2017}
{Luo}, B., {Brandt}, W.~N., {Xue}, Y.~Q., {et~al.} 2017, \apjs, 228, 2,
  \dodoi{10.3847/1538-4365/228/1/2}

\bibitem[{{Marchesi} {et~al.}(2016){Marchesi}, {Lanzuisi}, {Civano}, {Iwasawa},
  {Suh}, {Comastri}, {Zamorani}, {Allevato}, {Griffiths}, {Miyaji}, {Ranalli},
  {Salvato}, {Schawinski}, {Silverman}, {Treister}, {Urry}, \&
  {Vignali}}]{Marchesi2016}
{Marchesi}, S., {Lanzuisi}, G., {Civano}, F., {et~al.} 2016, \apj, 830, 100,
  \dodoi{10.3847/0004-637X/830/2/100}

\bibitem[{{Merloni} {et~al.}(2012){Merloni}, {Predehl}, {Becker},
  {B{\"o}hringer}, {Boller}, {Brunner}, {Brusa}, {Dennerl}, {Freyberg},
  {Friedrich}, {Georgakakis}, {Haberl}, {Hasinger}, {Meidinger}, {Mohr},
  {Nandra}, {Rau}, {Reiprich}, {Robrade}, {Salvato}, {Santangelo}, {Sasaki},
  {Schwope}, {Wilms}, \& {German eROSITA Consortium}}]{Merloni2012}
{Merloni}, A., {Predehl}, P., {Becker}, W., {et~al.} 2012, arXiv e-prints,
  arXiv:1209.3114.
\newblock \doarXiv{1209.3114}

\bibitem[{{Nandra} {et~al.}(2005){Nandra}, {Laird}, {Adelberger}, {Gardner},
  {Mushotzky}, {Rhodes}, {Steidel}, {Teplitz}, \& {Arnaud}}]{Nandra2005}
{Nandra}, K., {Laird}, E.~S., {Adelberger}, K., {et~al.} 2005, \mnras, 356,
  568, \dodoi{10.1111/j.1365-2966.2004.08475.x}

\bibitem[{{Nandra} {et~al.}(2015){Nandra}, {Laird}, {Aird}, {Salvato},
  {Georgakakis}, {Barro}, {Perez-Gonzalez}, {Barmby}, {Chary}, {Coil},
  {Cooper}, {Davis}, {Dickinson}, {Faber}, {Fazio}, {Guhathakurta}, {Gwyn},
  {Hsu}, {Huang}, {Ivison}, {Koo}, {Newman}, {Rangel}, {Yamada}, \&
  {Willmer}}]{Nandra2015}
{Nandra}, K., {Laird}, E.~S., {Aird}, J.~A., {et~al.} 2015, \apjs, 220, 10,
  \dodoi{10.1088/0067-0049/220/1/10}

\bibitem[{{Pineau} {et~al.}(2017){Pineau}, {Derriere}, {Motch}, {Carrera},
  {Genova}, {Michel}, {Mingo}, {Mints}, {Nebot G{\'o}mez-Mor{\'a}n}, {Rosen},
  \& {Ruiz Camu{\~n}as}}]{Pineau2017}
{Pineau}, F.~X., {Derriere}, S., {Motch}, C., {et~al.} 2017, \aap, 597, A89,
  \dodoi{10.1051/0004-6361/201629219}

\bibitem[{{Ranalli} {et~al.}(2015){Ranalli}, {Georgantopoulos}, {Corral},
  {Koutoulidis}, {Rovilos}, {Carrera}, {Akylas}, {Del Moro}, {Georgakakis},
  {Gilli}, \& {Vignali}}]{Ranalli2015}
{Ranalli}, P., {Georgantopoulos}, I., {Corral}, A., {et~al.} 2015, \aap, 577,
  A121, \dodoi{10.1051/0004-6361/201425246}

\bibitem[{{Rosen} {et~al.}(2016){Rosen}, {Webb}, {Watson}, {Ballet}, {Barret},
  {Braito}, {Carrera}, {Ceballos}, {Coriat}, {Della Ceca}, {Denkinson},
  {Esquej}, {Farrell}, {Freyberg}, {Gris{\'e}}, {Guillout}, {Heil},
  {Koliopanos}, {Law-Green}, {Lamer}, {Lin}, {Martino}, {Michel}, {Motch},
  {Nebot Gomez-Moran}, {Page}, {Page}, {Page}, {Pakull}, {Pye}, {Read},
  {Rodriguez}, {Sakano}, {Saxton}, {Schwope}, {Scott}, {Sturm}, {Traulsen},
  {Yershov}, \& {Zolotukhin}}]{Rosen2016}
{Rosen}, S.~R., {Webb}, N.~A., {Watson}, M.~G., {et~al.} 2016, \aap, 590, A1,
  \dodoi{10.1051/0004-6361/201526416}

\bibitem[{{Salvato} {et~al.}(2018){Salvato}, {Buchner}, {Budav{\'a}ri},
  {Dwelly}, {Merloni}, {Brusa}, {Rau}, {Fotopoulou}, \& {Nand
  ra}}]{Salvato2018}
{Salvato}, M., {Buchner}, J., {Budav{\'a}ri}, T., {et~al.} 2018, \mnras, 473,
  4937, \dodoi{10.1093/mnras/stx2651}

\bibitem[{{Schlafly} {et~al.}(2019){Schlafly}, {Meisner}, \&
  {Green}}]{Schlafly2019}
{Schlafly}, E.~F., {Meisner}, A.~M., \& {Green}, G.~M. 2019, \apjs, 240, 30,
  \dodoi{10.3847/1538-4365/aafbea}

\bibitem[{{Shen} {et~al.}(2015){Shen}, {Brandt}, {Dawson}, {Hall}, {McGreer},
  {Anderson}, {Chen}, {Denney}, {Eftekharzadeh}, {Fan}, {Gao}, {Green},
  {Greene}, {Ho}, {Horne}, {Jiang}, {Kelly}, {Kinemuchi}, {Kochanek},
  {P{\^a}ris}, {Peters}, {Peterson}, {Petitjean}, {Ponder}, {Richards},
  {Schneider}, {Seth}, {Smith}, {Strauss}, {Tao}, {Trump}, {Wood-Vasey}, {Zu},
  {Eisenstein}, {Pan}, {Bizyaev}, {Malanushenko}, {Malanushenko}, \&
  {Oravetz}}]{Shen2015}
{Shen}, Y., {Brandt}, W.~N., {Dawson}, K.~S., {et~al.} 2015, \apjs, 216, 4,
  \dodoi{10.1088/0067-0049/216/1/4}

\bibitem[{{Shen} {et~al.}(2019){Shen}, {Hall}, {Horne}, {Zhu}, {McGreer},
  {Simm}, {Trump}, {Kinemuchi}, {Brandt}, {Green}, {Grier}, {Guo}, {Ho},
  {Homayouni}, {Jiang}, {I-Hsiu Li}, {Morganson}, {Petitjean}, {Richards},
  {Schneider}, {Starkey}, {Wang}, {Chambers}, {Kaiser}, {Kudritzki}, {Magnier},
  \& {Waters}}]{Shen2019}
{Shen}, Y., {Hall}, P.~B., {Horne}, K., {et~al.} 2019, \apjs, 241, 34,
  \dodoi{10.3847/1538-4365/ab074f}

\bibitem[{{Smee} {et~al.}(2013){Smee}, {Gunn}, {Uomoto}, {Roe}, {Schlegel},
  {Rockosi}, {Carr}, {Leger}, {Dawson}, {Olmstead}, {Brinkmann}, {Owen},
  {Barkhouser}, {Honscheid}, {Harding}, {Long}, {Lupton}, {Loomis}, {Anderson},
  {Annis}, {Bernardi}, {Bhardwaj}, {Bizyaev}, {Bolton}, {Brewington}, {Briggs},
  {Burles}, {Burns}, {Castander}, {Connolly}, {Davenport}, {Ebelke}, {Epps},
  {Feldman}, {Friedman}, {Frieman}, {Heckman}, {Hull}, {Knapp}, {Lawrence},
  {Loveday}, {Mannery}, {Malanushenko}, {Malanushenko}, {Merrelli}, {Muna},
  {Newman}, {Nichol}, {Oravetz}, {Pan}, {Pope}, {Ricketts}, {Shelden},
  {Sandford}, {Siegmund}, {Simmons}, {Smith}, {Snedden}, {Schneider},
  {SubbaRao}, {Tremonti}, {Waddell}, \& {York}}]{Smee2013}
{Smee}, S.~A., {Gunn}, J.~E., {Uomoto}, A., {et~al.} 2013, \aj, 146, 32,
  \dodoi{10.1088/0004-6256/146/2/32}

\bibitem[{{Tonry} {et~al.}(2012){Tonry}, {Stubbs}, {Kilic}, {Flewelling},
  {Deacon}, {Chornock}, {Berger}, {Burgett}, {Chambers}, {Kaiser}, {Kudritzki},
  {Hodapp}, {Magnier}, {Morgan}, {Price}, \& {Wainscoat}}]{Tonry2012}
{Tonry}, J.~L., {Stubbs}, C.~W., {Kilic}, M., {et~al.} 2012, \apj, 745, 42,
  \dodoi{10.1088/0004-637X/745/1/42}

\bibitem[{{Watson} {et~al.}(2009){Watson}, {Schr{\"o}der}, {Fyfe}, {Page},
  {Lamer}, {Mateos}, {Pye}, {Sakano}, {Rosen}, {Ballet}, {Barcons}, {Barret},
  {Boller}, {Brunner}, {Brusa}, {Caccianiga}, {Carrera}, {Ceballos}, {Della
  Ceca}, {Denby}, {Denkinson}, {Dupuy}, {Farrell}, {Fraschetti}, {Freyberg},
  {Guillout}, {Hambaryan}, {Maccacaro}, {Mathiesen}, {McMahon}, {Michel},
  {Motch}, {Osborne}, {Page}, {Pakull}, {Pietsch}, {Saxton}, {Schwope},
  {Severgnini}, {Simpson}, {Sironi}, {Stewart}, {Stewart}, {Stobbart}, {Tedds},
  {Warwick}, {Webb}, {West}, {Worrall}, \& {Yuan}}]{Watson2009}
{Watson}, M.~G., {Schr{\"o}der}, A.~C., {Fyfe}, D., {et~al.} 2009, \aap, 493,
  339, \dodoi{10.1051/0004-6361:200810534}

\bibitem[{{Wen} {et~al.}(2012){Wen}, {Han}, \& {Liu}}]{Wen2012}
{Wen}, Z.~L., {Han}, J.~L., \& {Liu}, F.~S. 2012, \apjs, 199, 34,
  \dodoi{10.1088/0067-0049/199/2/34}

\bibitem[{{White} {et~al.}(1997){White}, {Becker}, {Helfand}, \&
  {Gregg}}]{White1997}
{White}, R.~L., {Becker}, R.~H., {Helfand}, D.~J., \& {Gregg}, M.~D. 1997,
  \apj, 475, 479, \dodoi{10.1086/303564}

\bibitem[{{Wright} {et~al.}(2010){Wright}, {Eisenhardt}, {Mainzer}, {Ressler},
  {Cutri}, {Jarrett}, {Kirkpatrick}, {Padgett}, {McMillan}, {Skrutskie},
  {Stanford}, {Cohen}, {Walker}, {Mather}, {Leisawitz}, {Gautier}, {McLean},
  {Benford}, {Lonsdale}, {Blain}, {Mendez}, {Irace}, {Duval}, {Liu}, {Royer},
  {Heinrichsen}, {Howard}, {Shannon}, {Kendall}, {Walsh}, {Larsen}, {Cardon},
  {Schick}, {Schwalm}, {Abid}, {Fabinsky}, {Naes}, \& {Tsai}}]{Wright2010}
{Wright}, E.~L., {Eisenhardt}, P. R.~M., {Mainzer}, A.~K., {et~al.} 2010, \aj,
  140, 1868, \dodoi{10.1088/0004-6256/140/6/1868}

\bibitem[{{York} {et~al.}(2000){York}, {Adelman}, {Anderson}, {Anderson},
  {Annis}, {Bahcall}, {Bakken}, {Barkhouser}, {Bastian}, {Berman}, {Boroski},
  {Bracker}, {Briegel}, {Briggs}, {Brinkmann}, {Brunner}, {Burles}, {Carey},
  {Carr}, {Castander}, {Chen}, {Colestock}, {Connolly}, {Crocker}, {Csabai},
  {Czarapata}, {Davis}, {Doi}, {Dombeck}, {Eisenstein}, {Ellman}, {Elms},
  {Evans}, {Fan}, {Federwitz}, {Fiscelli}, {Friedman}, {Frieman}, {Fukugita},
  {Gillespie}, {Gunn}, {Gurbani}, {de Haas}, {Haldeman}, {Harris}, {Hayes},
  {Heckman}, {Hennessy}, {Hindsley}, {Holm}, {Holmgren}, {Huang}, {Hull},
  {Husby}, {Ichikawa}, {Ichikawa}, {Ivezi{\'c}}, {Kent}, {Kim}, {Kinney},
  {Klaene}, {Kleinman}, {Kleinman}, {Knapp}, {Korienek}, {Kron}, {Kunszt},
  {Lamb}, {Lee}, {Leger}, {Limmongkol}, {Lindenmeyer}, {Long}, {Loomis},
  {Loveday}, {Lucinio}, {Lupton}, {MacKinnon}, {Mannery}, {Mantsch}, {Margon},
  {McGehee}, {McKay}, {Meiksin}, {Merelli}, {Monet}, {Munn}, {Narayanan},
  {Nash}, {Neilsen}, {Neswold}, {Newberg}, {Nichol}, {Nicinski}, {Nonino},
  {Okada}, {Okamura}, {Ostriker}, {Owen}, {Pauls}, {Peoples}, {Peterson},
  {Petravick}, {Pier}, {Pope}, {Pordes}, {Prosapio}, {Rechenmacher}, {Quinn},
  {Richards}, {Richmond}, {Rivetta}, {Rockosi}, {Ruthmansdorfer}, {Sand ford},
  {Schlegel}, {Schneider}, {Sekiguchi}, {Sergey}, {Shimasaku}, {Siegmund},
  {Smee}, {Smith}, {Snedden}, {Stone}, {Stoughton}, {Strauss}, {Stubbs},
  {SubbaRao}, {Szalay}, {Szapudi}, {Szokoly}, {Thakar}, {Tremonti}, {Tucker},
  {Uomoto}, {Vanden Berk}, {Vogeley}, {Waddell}, {Wang}, {Watanabe},
  {Weinberg}, {Yanny}, {Yasuda}, \& {SDSS Collaboration}}]{York2000}
{York}, D.~G., {Adelman}, J., {Anderson}, John~E., J., {et~al.} 2000, \aj, 120,
  1579, \dodoi{10.1086/301513}

\end{thebibliography}
\end{CJK*}
\end{document}